\documentclass[preprint2]{aastex63}

\accepted{June 7, 2021}
\submitjournal{ApJ}

\shorttitle{A Giant Loop in Tumultuous IC 5063}
\shortauthors{Maksym et al.}

\usepackage[applemac]{inputenc}
\usepackage{hyperref,breakurl}
\def\UrlBreaks{\do\/\do-}
\usepackage{ifthen,graphicx}
\usepackage{epstopdf,placeins}
\usepackage{soul,color,amsmath}
\usepackage{epstopdf,overpic}
\usepackage{soul,color,amsmath}
\def\mathnew{\mathsurround=0pt}
\def\simov#1#2{\lower 2.5pt\vbox{\baselineskip0pt \lineskip-.5pt
\ialign{$\mathnew#1\hfil##\hfil$\crcr#2\crcr\sim\crcr}}}
\def\simless{\mathrel{\mathpalette\simov <}}
\def\simgreat{\mathrel{\mathpalette\simov >}}
\newcommand{\MeV}{Me\kern-0.11em V}
\newcommand{\keV}{ke\kern-0.11em V}

\newcommand{\z}{{\it z\/}}

\newcommand{\ecmss}{erg~cm$^{-2}$ s$^{-1}$}
\newcommand{\es}{\ensuremath{\rm erg~s}^{-1}}
\newcommand{\kms}{\ensuremath{\rm km\,s}^{-1}}

\newcommand{\icgsbri}{\ensuremath{\rm{erg\;cm}^{-2}\;\rm{s}^{-1}\rm{arcsec}^{-2}}}

\newcommand{\Mbh}{\ensuremath{M_{\bullet}}}
 
\newcommand{\Msun}{\ensuremath{{\rm M}_{\odot}}}

\newcommand{\Lbol}{\ensuremath{L_{bol}}}

\makeatletter
\newcommand\iont[2]{{#1$\;${\small\expandafter\@slowromancap\romannumeral #2@\relax}}}
\makeatother
\makeatletter
\newcommand{\raisemath}[1]{\mathpalette{\raisem@th{#1}}}
\newcommand{\raisem@th}[3]{\raisebox{#1}{$#2#3$}}
\makeatother

\begin{document}

\title{A Giant Loop of Ionized Gas Emerging from \\ the Tumultuous Central Region of IC 5063\footnote{Based on observations made with the NASA/ESA Hubble Space Telescope, obtained from the data archive at the Space Telescope Science Institute. STScI is operated by the Association of Universities for Research in Astronomy, Inc. under NASA contract NAS 5-26555. These observations are associated with programs \#8598 and \#15609.}}

\correspondingauthor{W. Peter Maksym; @StellarBones}
\email{walter.maksym@cfa.harvard.edu}

\author[0000-0002-2203-7889]{W. Peter Maksym}
\affiliation{Center for Astrophysics \textbar\ Harvard  \& Smithsonian, 60 Garden St., Cambridge, MA 02138, USA}

\author[0000-0002-3554-3318]{Giuseppina Fabbiano}
\affiliation{Center for Astrophysics \textbar\ Harvard  \& Smithsonian, 60 Garden St., Cambridge, MA 02138, USA}

\author[0000-0001-5060-1398]{Martin Elvis}
\affiliation{Center for Astrophysics \textbar\ Harvard  \& Smithsonian, 60 Garden St., Cambridge, MA 02138, USA}

\author[0000-0001-6947-5846]{Luis C. Ho}
\affiliation{Kavli Institute for Astronomy and Astrophysics, Peking University, Beijing 100871, People’s Republic of China}
\affiliation{Department of Astronomy, School of Physics, Peking University, Beijing 100871, People’s Republic of China}

\author[0000-0002-0616-6971]{Tom Oosterloo}
\affiliation{ASTRON, The Netherlands Institute for Radio Astronomy, Postbus 2, 7990 AA Dwingeloo, The Netherlands}
\affiliation{Kapteyn Astronomical Institute, University of Groningen, Postbus 800, 9700 AV Groningen, The Netherlands }

\author[0000-0003-4178-0800]{Jingzhe Ma}
\affiliation{Center for Astrophysics \textbar\ Harvard \& Smithsonian, 60 Garden St., Cambridge, MA 02138, USA}

\author{Andrea Travascio}
\affiliation{INAF-Osservatorio Astronomico di Trieste, via G.B. Tiepolo 11, I-34131, Trieste, Italy}

\author[0000-0002-3365-8875]{Travis C. Fischer}
\affiliation{Space Telescope Science Institute,
3700 San Martin Drive, Baltimore, MD 21218, USA}

\author[0000-0002-6131-9539]{William C. Keel}
\affiliation{Department of Physics and Astronomy, University of Alabama, Box 870324, Tuscaloosa, AL 35487, USA}



\begin{abstract}

The biconical radiation pattern extending from an active galactic nucleus (AGN) may strongly photoionize the circumnuclear interstellar medium (ISM) and stimulate emission from the narrow line region (NLR).  Observations of the NLR may provide clues to the structure of dense material that preferentially obscures the bicone at certain angles, and may reveal the presence of processes in the ISM tied to AGN accretion and feedback.  Ground-based integral field units (IFUs) may study these processes via well-understood forbidden diagnostic lines such as [\ion{O}{3}] and [\ion{S}{2}], but scales of $\sim10$s of pc remain challenging to spatially resolve at these wavelengths for all but the nearest AGN.  We present recent narrow filter {\it Hubble Space Telescope} ({\it HST}) observations of diagnostic forbidden ([\ion{O}{3}], [\ion{S}{2}]) and Balmer (H$\alpha$, H$\beta$) lines in the NLR of IC 5063.  This AGN's jet inclination into the plane of the galaxy provides an important laboratory for strong AGN-host interactions.  We find evidence for a low-ionization loop which emits brightly in [\ion{S}{2}] and [\ion{N}{2}], and which may arise from plume-like hot outflows that ablate ISM from the galactic plane before escaping laterally.  We also present spatially resolved Baldwin-Phillips-Terlevich diagnostic maps of the IC 5063 NLR.  These maps suggest a sharp transition to lower-ionization states outside the jet path, and that such emission is dominated by $\sim10-40$\,pc clumps and filamentary structure at large ($>>25\degr$) angles from the bicone axis.  Such emission may arise from precursorless shocks when AGN outflows impact low-density hot plasma in the cross-cone.

\end{abstract}

\keywords{galaxies: active --- 
galaxies: individual (IC 5063) --- galaxies: Seyfert}


\section{Introduction} \label{sec:intro}

IC 5063 is a nearby ($z=0.01140$;  47.9 Mpc) galaxy that hosts a highly obscured (log\,$[n_{\rm H}/{\rm cm}^{-2}]=23.55$; \citealt{Ricci17}) and moderately powerful ($\Lbol\sim7.7\times10^{44}\,\es$, $\Mbh\sim2.8\times10^8\Msun$, $\eta_{Edd}=\Lbol/(\Mbh(L_{Edd}/\Msun))$; see  \citealt{Nicastro03,Morganti07}) active galactic nucleus (AGN). 
The brightest extended narrow line (ENLR) emission is associated with powerful kpc-scale radio outflows in the X-rays \citep{GG17} and optical \citep{Danziger81,Morganti98,Schmitt03,Mingozzi19}, but fainter ENLR emission is detectable out to $\simgreat10\,$kpc.  Notably, the outflows are oriented directly into the plane of the galaxy, colliding with the nuclear interstellar medium (ISM).  Models of these jet-ISM interactions by \cite{Mukherjee18} broadly replicate the key CO features found by \cite{Morganti15} with ALMA.  These models also predict venting of hot plasma perpendicular to the disk and entrainment of cooler clumps and filaments in the halo.

\cite{Maksym20} recently described spectacular large-scale ($\simgreat11$\,kpc) dark and light ``rays" which IC 5063 displays in {\it HST} continuum imaging at large angles ($\simgreat60\degr$) relative to both the galactic plane and jet, extending from the nucleus. These ``rays" were visually discovered by J. Schmidt\footnote{\url{https://twitter.com/SpaceGeck/status/1201350966945017856}} in {\it HST} continuum imaging (ACS F814W) and confirmed by \cite{Maksym20} as azimuthal minima in the best stellar brightness profile fits to both ACS F814W and WFC3 F763M imaging.

\cite{Maksym20} suggested that the continuum excess (i.e. the brighter counterpart to the azimuthal minima) was unlikely to result from stellar structure as might arise from blending of highly evolved stellar shells after a minor galactic merger.  Rather, this continuum excess was likely the result of reflection of AGN continuum emission by a halo of diffuse dust extended over galactic scales, at azimuthal angles as large as $\sim70\degr$ from the bicone axis.  They suggested that the dark ``rays" might be formed through shadows cast from the AGN (such as by dense dust and gas in torus), or through removal of the reflective dust at some angles by hot lateral outflows \citep[again, as in][]{Mukherjee18}.

Current AGN models and observations \citep{Elvis00, HA18} typically entail wavelength-dependent stratification of radiative transmission as a function of the angle from the nuclear axis, with highly-ionizing radiation more readily exciting low-density material near the center of the bicone.  In \cite{Maksym16}, we used {\it HST} narrow filter observations of the NLR to map the ionization structure of the nearby ($D=53$\,Mpc) Seyfert 2 galaxy NGC 3393 via the commonly-used BPT diagnostic (Baldwin-Phillips-Terlevich, \citealt{BPT81}; see also \citealt{VO87}) at $\sim10$\,pc scales.  Integrated field units (IFUs) are commonly used for such mapping via spatially resolved spectroscopy \citep[e.g.][]{Mingozzi19}, but even IFUs with adaptive optics depend upon exceptional seeing to approach {\it HST} resolution, with additional challenges near the rest-frame wavelengths of e.g. H$\beta$.

In this paper, we study the inner $\sim$kpc of IC 5063 using techniques similar to the \cite{Calzetti04} study of 4 starburst galaxies, and to NGC 3393 in \cite{Maksym16}.  Unlike \cite{Maksym16}, we use a full complement of [\ion{O}{3}]\,$\lambda5007$\,\AA, H$\alpha$\,$\lambda6563\,$\AA, H$\beta$\,$\lambda4861\,$\AA, and [\ion{S}{2}] $\lambda\lambda6716,6731\,$\AA\AA.   With direct measurements of H$\beta$, there is no need to infer it from H$\alpha$ (which would produce inaccurate values contaminated by substantial dust).  As with NGC 3393, we identify geometrically thin ($\sim$pixel-scale), LINER-like structure (LINER: low ionization nuclear emission-line region) which covers the edges of bicone and would be lost to observations with worse angular resolution.  We also identify a forbidden-line projected loop near the nucleus which may indicate venting of hot material ablated from the inner disk, as described by \cite{Mukherjee18}.

As in \cite{Oosterloo17}, we adopt an angular size distance of 47.9 Mpc and a scale of 1\arcsec = 232 pc for IC 5063.

\begin{deluxetable*}{ccccccc}
\tablecaption{Hubble Observation Properties}
\tablehead{
\colhead{Program ID} & \colhead{PI} & \colhead{Obs Date}  & \colhead{Exposure (s)} & \colhead{Instrument} & \colhead{Filter} & \colhead{Note}
}
\label{table:hstobs}
\startdata
\tableline
\tableline
8598	& Schmitt & 	2001 Apr 6	& 600	&	WFPC2	&  FR533N &  [\ion{O}{3}]		\\
8598	& Schmitt &	2001 Apr 6	& 80	&	WFPC2	&  F547M &  blue continuum		\\
\tableline
15609	& Fabbiano&	2019 Mar 7	& 1012	&	WFC3/UVIS	&  F665N & H$\alpha$+[\ion{N}{2}]	\\
15609	& Fabbiano&	2019 Mar 7	& 1612	&	WFC3/UVIS	&  F673N & [\ion{S}{2}]	\\
15609	& Fabbiano&	2019 Mar 7	& 304	&	WFC3/UVIS	&  F763M & red continuum	\\
15609	& Fabbiano&	2019 Mar 8	& 2166	&	WFC3/UVIS	&  FQ387N & H$\beta$	\\
\tableline
\enddata
\end{deluxetable*}

\section{Observations and Data}

\subsection{HST Observations}

IC 5063 has been observed extensively by {\it HST} over its mission using multiple filters of WFPC2, ACS and NICMOS.  The archival dataset includes 600s of WFPC2 with ramp filter FR533N set to $\lambda=5064\,$\AA\ to observe [\ion{O}{3}] emission at $\lambda_{rest}=5007\,$\AA, as well as  an adjacent continuum band F547M (PI Schmitt, Program ID 8598).   We have also recently observed IC 5063 as part of a program to map the nucleus according to standard AGN diagnostics (on a 2-D plane with [\ion{O}{3}]/H$\beta$ vs. [\ion{S}{2}]/H$\alpha$; see \citealt{BPT81,VO87}) at scales of $\sim20\,$pc \citep[Program ID 15609, PI Fabbiano; see][]{Maksym16}.  These observations are summarized in Table \ref{table:hstobs}.  All instrumental conversions to flux are inferred from standard {\it HST} filter values, except where further modeled to isolate a spectral feature.

We generate BPT maps using the BPT prescription of \cite{Kewley06} to categorize each image pixel according to line ratios ([\ion{O}{3}]/H$\beta$,  [\ion{S}{2}]/H$\alpha$) as Seyfert-like, LINER-like and star-forming (SF).  In order to do this, we used methods similar to \cite{Maksym16} to reduce the narrow-band images and continuua for  [\ion{O}{3}]\,$\lambda5007$\,\AA, H$\alpha$\,$\lambda6563\,$\AA, H$\beta$\,$\lambda4861\,$\AA, and [\ion{S}{2}] $\lambda\lambda6716,6731\,$\AA\AA.  We use the DR2 catalog from \cite{GaiaDR2}  for astrometric cross-registration before re-sampling the images to a common pixel grid with {\tt AstroDrizzle}.  

For line ratio images, we sample at a pixel scale of $0\farcs09$, which is slightly less than the WFPC2 native pixel size $\sim0\farcs1$.  This sets our effective angular resolution limit (undersampling the PSF), and improves WFC3 signal-to-noise in fainter regions of interest.  For comparison, the FWHM$\sim0\farcs07$ for the WFC3 PSF.  We also sample higher-resolution WFC3 images for inspection ($0\farcs046$).  We subtract a sky value from an object-free region and convert count rate images to flux images using the PHOTFLAM header keyword.  We then subtract the continuum from emission line images iteratively rescaling the continuum image to minimize residuals in those regions dominated by the stellar continuum.  First, we identify off-nuclear regions with strong continuum emission and little evidence of dust or line emission, particularly in the F547M continuum band where we expect dust absorption to be more pronounced.  We then introduce a scaling factor to the continuum band such that the local median values from the continuum filter and the narrow filter are identical.  We model the [\ion{O}{3}] contamination of the F547M continuum assuming a 3:1 ratio for [\ion{O}{3}]\,$\lambda5007,4959$\,\AA.

Dust remains a problem in IC 5063, even for F763M as a continuum band.  The F763M/F814W map in \cite{Maksym20} Fig. 1 shows strong reddening even for these bands, which overlap between  $\lambda7165\,$\AA\ and $\lambda8092\,$\AA\ (observer frame).
The areas most strongly contaminated by dust are subject to poor continuum subtraction, with localized oversubtraction of H$\beta$ and H$\alpha$.  In order to correct for this effect, we assume that the color of the low-dust region used for rescaling is representative for the nuclear stellar population.  We then introduce a wavelength-dependent reddening correction to each filter according to \cite{Calzetti00}.  

Contamination of H$\alpha$ by the adjacent [\ion{N}{2}] lines ($\lambda\lambda6548,6584\,$\AA\AA) is more challenging to model, but like \cite{Calzetti04} we can infer [\ion{N}{2}] from our [\ion{S}{2}] measurements, and from plausible assumptions relating to [\ion{N}{2}] emission (we assume a 1:3.06 ratio between [\ion{N}{2}]$\lambda6548\,$\AA\ and [\ion{N}{2}]$\lambda6584\,$\AA\ where [\ion{N}{2}]$\lambda6548\,$\AA\ is not available, \citealt {OF06}).  Since [\ion{N}{2}] and  [\ion{S}{2}] are both forbidden lines that trace similar low-ionization phenomena and have similar ionization energies, this may be preferable to \cite{Maksym16} and \cite{Ma21}, who assume a constant [\ion{N}{2}]/H$\alpha$ ratio.  

\cite{Koss17} measure [\ion{N}{2}]/[\ion{S}{2}]$=1.62$ using the South African Astronomical Observatory 1.9m telescope 1.9m with 5.00 \AA\ resolution and a 2\arcsec\ slit on the nucleus (including the dust lane).  Our region of interest covers a $\sim10\arcsec\times10\arcsec$ region, however, and we hope to spatially resolve diverse physical processes for which  [\ion{N}{2}]/[\ion{S}{2}] may not be uniform.  We therefore consider a range of physically plausible values.  

If all of the gas were emitted from a collisional shocks with a photoionized precursor (i.e. preshock gas exists that can be ionized by photons from hot plasma in the shock), \cite{Allen08} predict 1.33$<$[\ion{N}{2}]/[\ion{S}{2}]$<$2.16 for models spanning 100-1000\,km\,s$^{-1}$, with the smallest values corresponding to the fastest shocks.   [\ion{N}{2}] and [\ion{S}{2}] have common BPT trends in photoionization models relative to H$\alpha$, such that parallel evolution along isocontours to the \cite{Kewley06} extreme star formation (SF) lines corresponds roughly to changes in ionization parameter, and normal evolution corresponds to changes in spectral index \citep{Feltre16}.  If we solve  [\ion{N}{2}] for [\ion{S}{2}] with [\ion{O}{3}]/H$\beta$ as the common parameter along the SF line, we find 1.88$<$[\ion{N}{2}]/[\ion{S}{2}]$<$2.07.  \cite{Oh17} found that [\ion{N}{2}]/[\ion{S}{2}] evolves gradually as a function of the AGN Eddington fraction.  Assuming $\Mbh\sim2.8\times10^8\Msun$ and $\Lbol\sim7.7\times10^{44}\,\es$ \citep{Nicastro03,Morganti07}, \cite{Oh17} implies [\ion{N}{2}]/[\ion{S}{2}]=2.09.  The low [\ion{N}{2}]/[\ion{S}{2}]=1.62 nuclear measurement \citep{Koss17} therefore suggests that much of the circumnuclear emission could be either collisionally stimulated or strongly absorbed by nuclear the dust lane.

\begin{figure} 
\noindent
\begin{overpic}[width=0.98\linewidth,trim=0 0 1.5cm 0, clip]{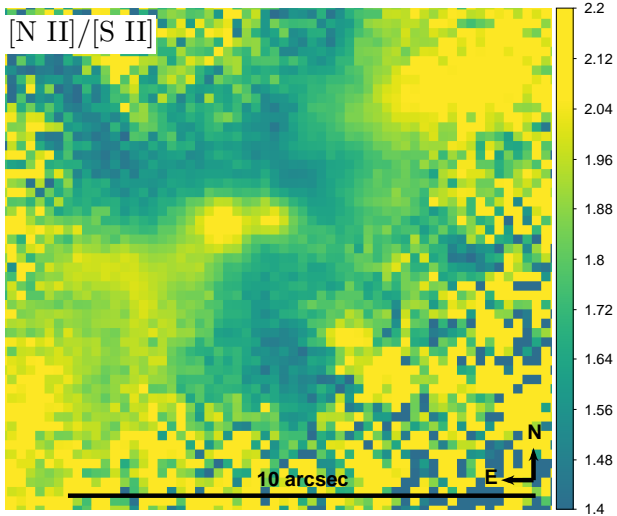}
	\put(2,75){\color{white}\rule{2.05cm}{0.5cm}}
	\put(3,77){{\parbox{2cm}{%
			[N II]/[S II]
			}}}
\end{overpic}\\
\caption{[\ion{N}{2}]/[\ion{S}{2}] map derived from MUSE integral field spectroscopy, taking the ratio between line pairs.  Typical values are $\sim1.8$, but extrema range between 1.4,2.2.} 
\label{fig:musen2s2}
\end{figure}

The integral field spectroscopy taken as science verification for MUSE on VLT in wide field mode \citep[][ESO program 60.A-9339, 2400s exposure on 2014 June 23rd]{Mingozzi19} only has $\sim0\farcs8$ angular resolution, but is capable of mapping [\ion{N}{2}]/[\ion{S}{2}] trends over the region of interest.  Using {\tt QFitsView} \footnote{\href{https://www.mpe.mpg.de/~ott/QFitsView/}{https://www.mpe.mpg.de/$\sim$ott/QFitsView/}}, we generated flux images by co-adding continuum-subtracted wavelength slices covering [\ion{N}{2}]$\lambda6584\,$\AA\ and [\ion{S}{2}] $\lambda\lambda6716,6731\,$\AA\AA.  We calculate [\ion{N}{2}]$\lambda6548\,$\AA\ from  [\ion{N}{2}]$\lambda6584\,$\AA\  to avoid contamination from H$\alpha$, which is closely blended and much brighter.    For $r<6\arcsec$, we find [\ion{N}{2}]/[\ion{S}{2}]$=1.85\pm{0.34}$, which corresponds well to the range of ratios under consideration.  We consider the median [\ion{N}{2}]/[\ion{S}{2}]=1.81 to be representative, and use it for subsequent calculations unless otherwise specified.  But as a check, we also investigate equally-spaced values of H$\alpha$ derived across the full range of physically plausible [\ion{N}{2}]/[\ion{S}{2}] (1.35, 1.58, 1.82, 2.05).  
Metallicity can affect  [\ion{N}{2}]/H$\alpha$ \citep[e.g. Fig. 21 in the][shock models]{Allen08}, such that local excesses in metallicity will increase [\ion{N}{2}]/[\ion{S}{2}], decrease our inferred H$\alpha$, and increase [\ion{S}{2}]/H$\alpha$.  These effects are most pronounced at low abundances, but  [\ion{N}{2}]/H$\alpha$ variations between solar and supersolar abundances (as might be expected in the circumnuclear ISM) are more modest.  Since the range of [\ion{N}{2}]/[\ion{S}{2}] is empirically constrained by the MUSE data, a full treatment of such abundance effects on the data reduction is beyond the scope of this paper.  

When calculating ratios of emission line fluxes at individual pixels, we limit the data to SNR$>3\sigma$, as per \cite{Maksym16}.  Since this tends to exclude faint regions outside the bicone that have physically interesting properties which may be visible by eye, we produce additional emission line maps which adaptively smooth $1\sigma-3\sigma$ pixels to enhance $3\sigma$ features at larger scales.  We use {\tt dmimgadapt} from {\it ciao} \citep{ciao}, which respects the undefined pixel values which we have masked at SNR$<1\sigma$.  Although these very low-significance pixels may include useful data, they also risk contaminating the adaptively smoothed features with negative value pixels produced by over-subtraction of locally invalid continuum.

\begin{figure*} 
\noindent
\begin{overpic}[width=0.32\linewidth]{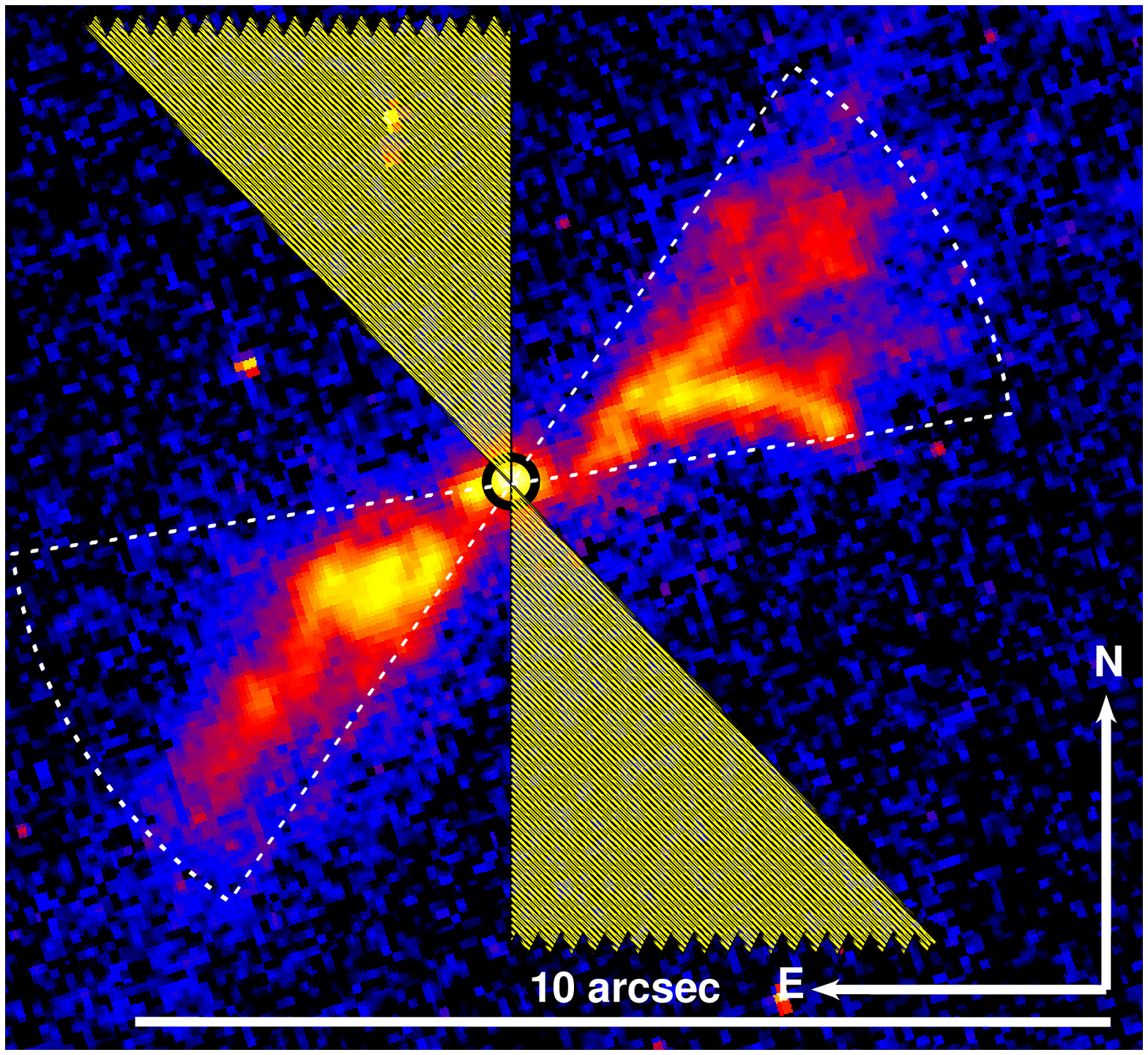}
	\put(5,75){\color{white}\rule{1.15cm}{0.5cm}}
	\put(-3,79){{\parbox{2cm}{%
			\begin{equation}
			\rm{[O\,III]} \nonumber
			\end{equation}
			}}}
\end{overpic}\hspace{0.001\textwidth}%
\begin{overpic}[width=0.32\linewidth]{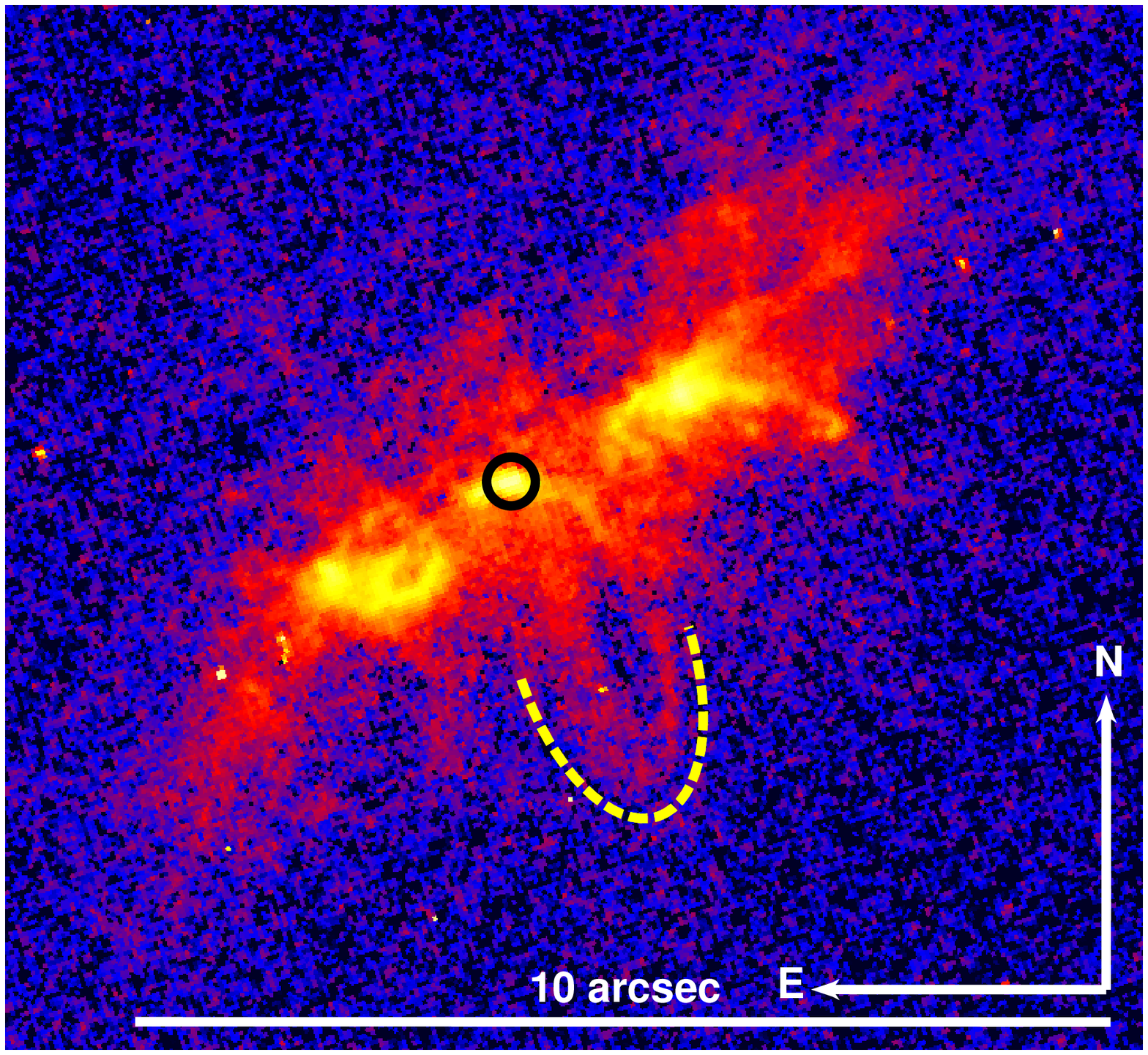}
	\put(7,75){\color{white}\rule{0.95cm}{0.5cm}}
	\put(-2,79){{\parbox{2cm}{%
			\begin{equation}
			\rm{[S\,II]} \nonumber
			\end{equation}
			}}}
\end{overpic}\hspace{0.001\textwidth}%
\begin{overpic}[width=0.32\linewidth]{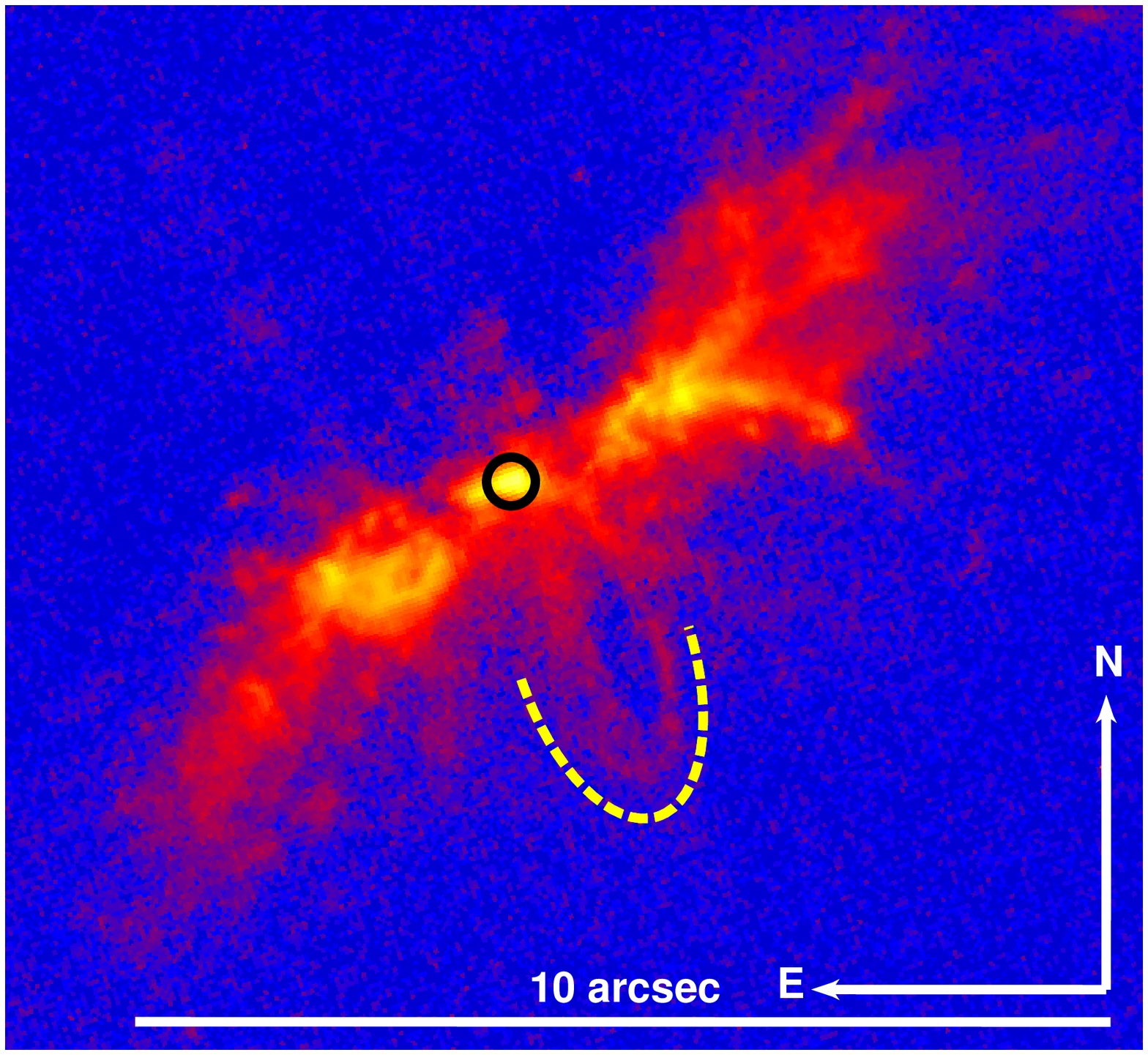}
	\put(3,75){\color{white}\rule{2.05cm}{0.5cm}}
	\put(4,79){{\parbox{2cm}{%
			\begin{equation}
			\rm{H}\alpha+\rm{[N\,II]} \nonumber
			\end{equation}
			}}}
\end{overpic}\\
\noindent
\begin{overpic}[width=0.32\linewidth]{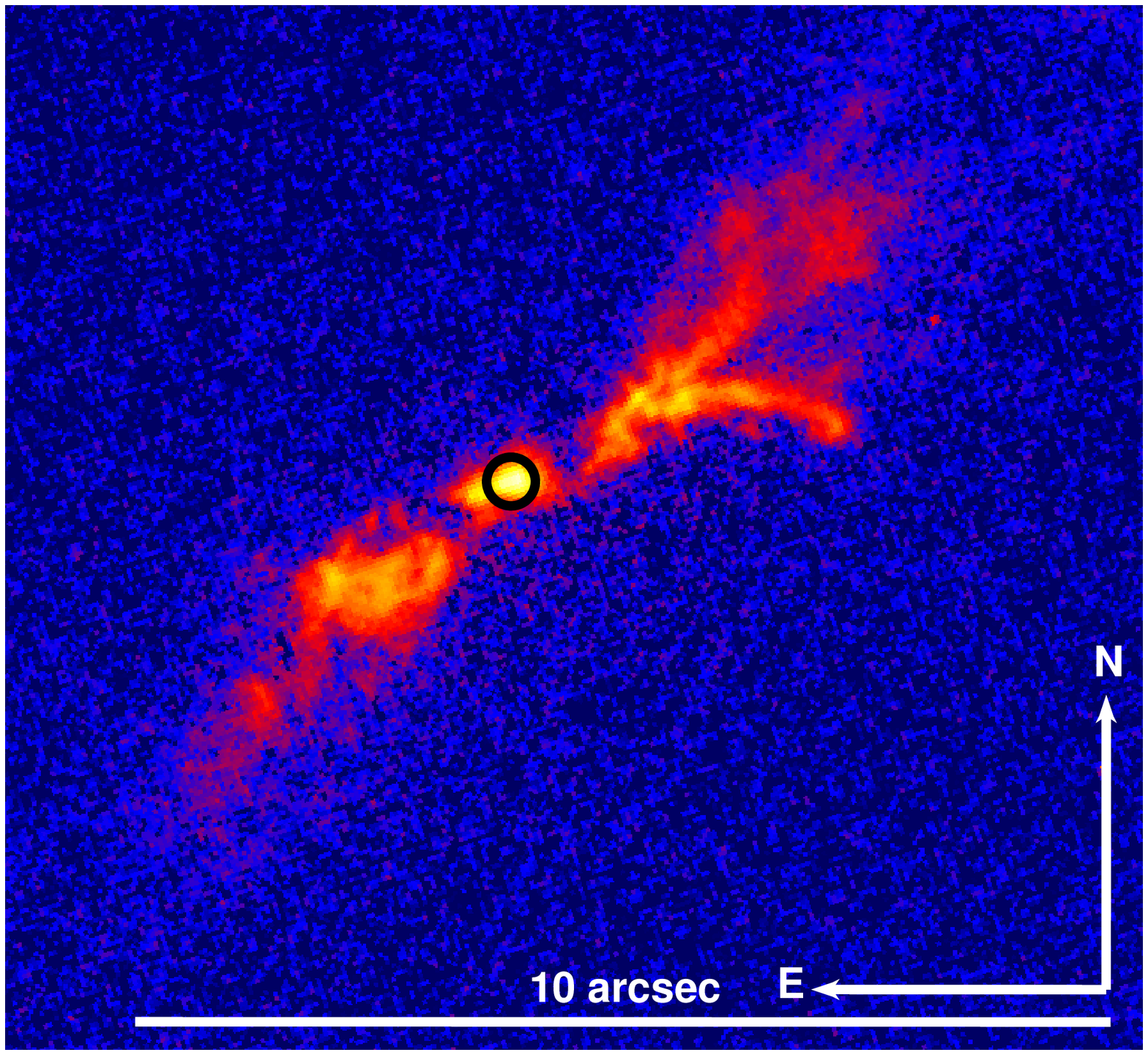}
	\put(3,75){\color{white}\rule{1.15cm}{0.5cm}}
	\put(-5,79){{\parbox{2cm}{%
			\begin{equation}
			\rm{H}\alpha \nonumber
			\end{equation}
			}}}
\end{overpic}\hspace{0.001\textwidth}%
\begin{overpic}[width=0.32\linewidth]{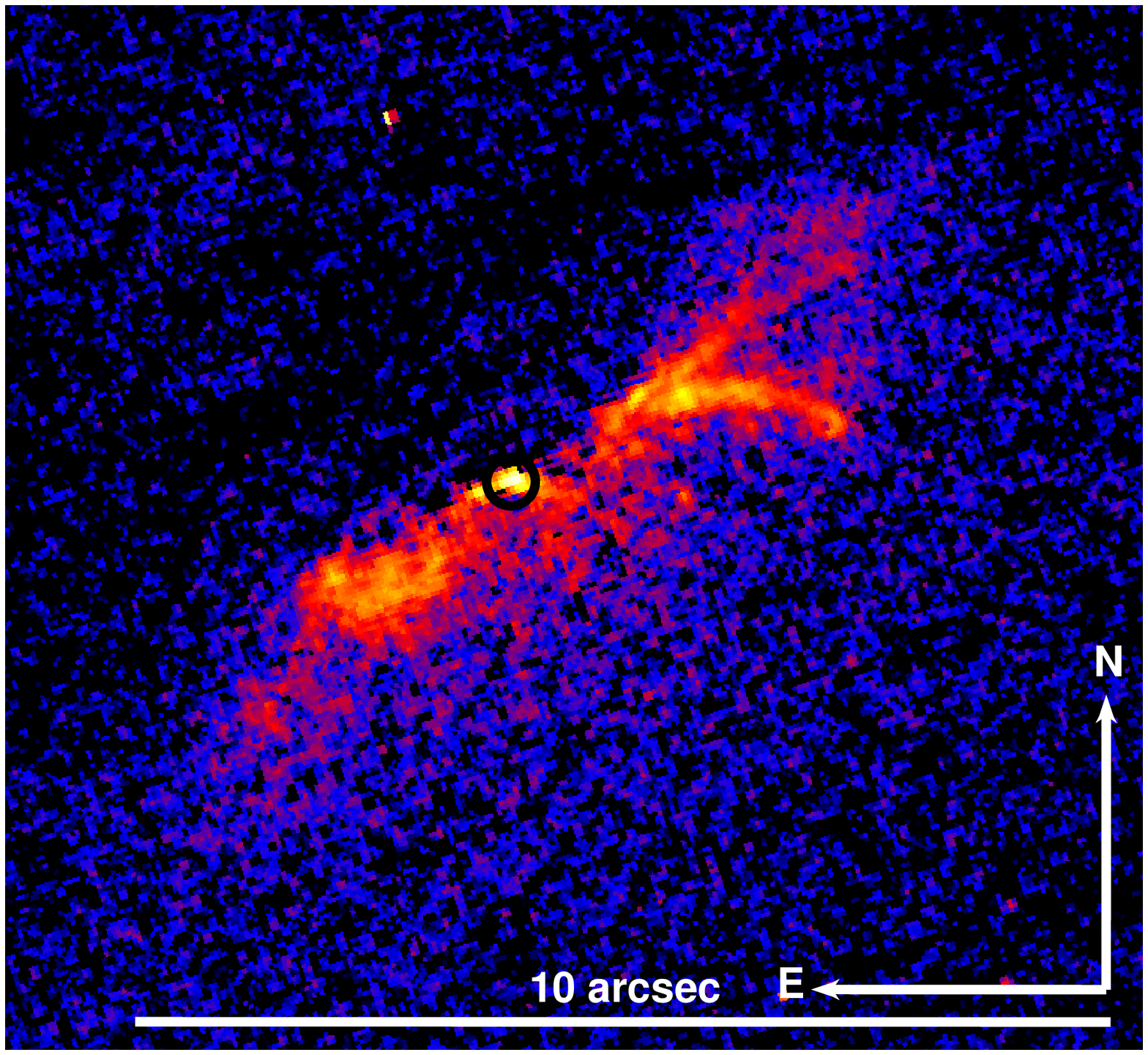}
	\put(3,75){\color{white}\rule{1.15cm}{0.5cm}}
	\put(-5,79){{\parbox{2cm}{%
			\begin{equation}
			\rm{H}\beta \nonumber
			\end{equation}
			}}}
\end{overpic}\hspace{0.001\textwidth}%
\begin{overpic}[width=0.32\linewidth]{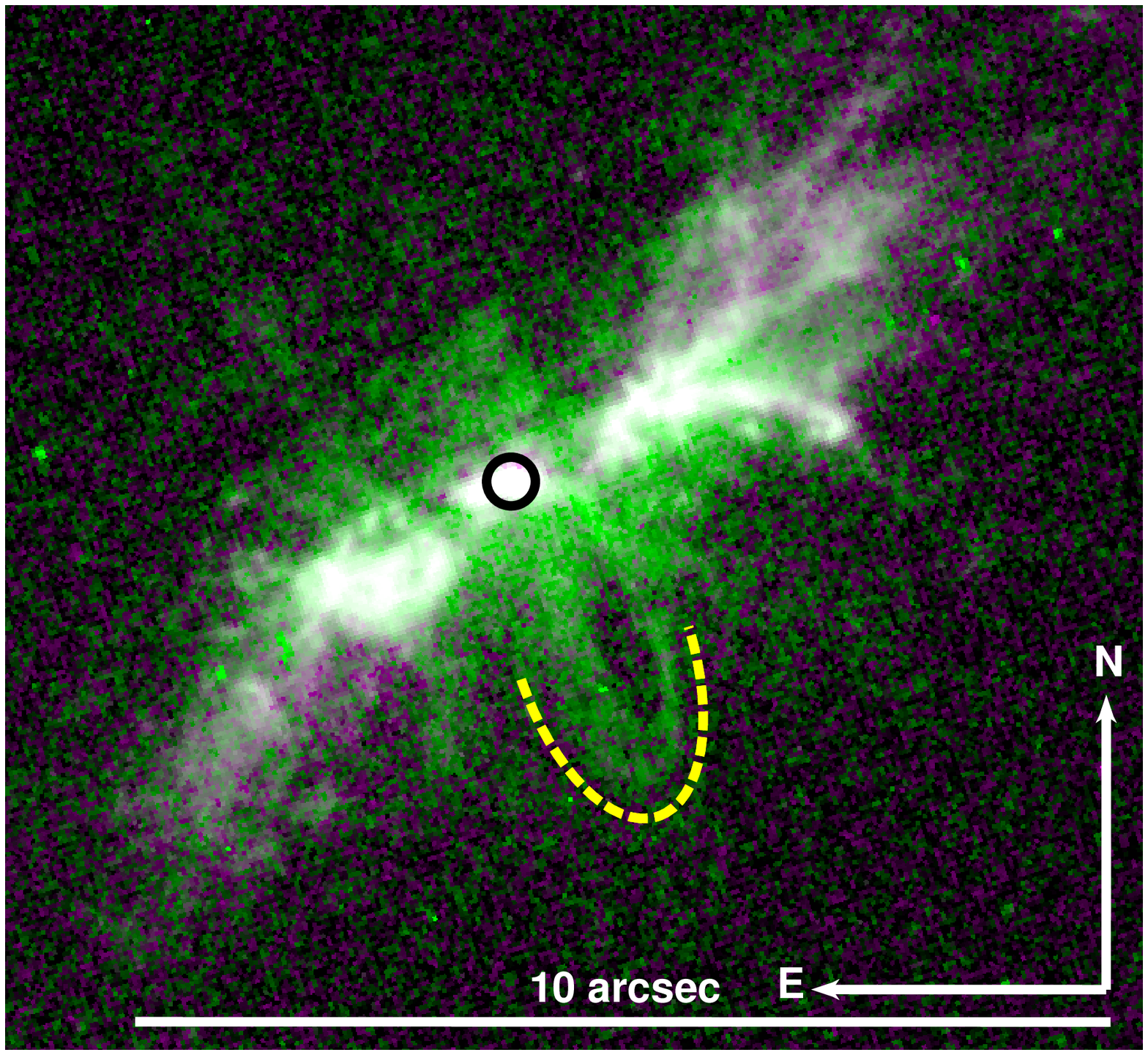}
	\multiput(27,50)(-1.8,3){10}{\color{yellow}\circle*{1}}
	\multiput(49,57)(-1.8,3){10}{\color{yellow}\circle*{1}}
	\put(29,23){\color{yellow}\circle{7}}
	\put(40,28){\color{yellow}\circle{10}}
	\put(65,43){\color{yellow}\circle{10}}
	\put(45,57){\color{yellow}\circle{7}}
	\put(43,62){\color{yellow}\circle{7}}
	\put(23,63){\color{yellow}\circle{10}}
	\put(1,75){\color{white}\rule{2.05cm}{0.5cm}}
	\put(2,79){{\parbox{2cm}{%
			\begin{equation}
			\rm{H}\alpha, \rm{[S\,II]} \nonumber
			\end{equation}
			}}}
\end{overpic}\\
\caption{{\bf Top:} Narrow line imaging of the IC 5063 nucleus and NLR for the filters which cover significant forbidden lines.  Stellar continuum imaging has been subtracted.  Scale and orientation are indicated, and wavebands are noted in the upper-left of each subfigure. The nucleus is marked with a black circle.  White dashed wedges indicate the NW and SE ionization cones.  Hashed shading marks the inner portion of the NE and SW dark ``rays" described by \citep{Maksym20}, which extend to $\simgreat50\arcsec$. {\bf Top-left:} [\ion{O}{3}] from archival WF/PC2 images sets the pixel scale. {\bf Top-center:} [\ion{S}{2}] tracks [\ion{O}{3}], but with an additional filamentary loop extending SW from the nucleus (indicated by a dashed yellow arc). {\bf Top-right:} The SW filamentary loop is also present in H$\alpha$+[\ion{N}{2}], which are spectrally confused by the F665N filter.  Pixels with at least one line below $3\sigma$ are black.  {\bf Bottom}: As above, but for Balmer lines. {\bf Bottom-left:} H$\alpha$ is spectrally confused with [\ion{N}{2}], so we model [\ion{N}{2}] contamination as a function of [\ion{S}{2}] using the median value from Fig. \ref{fig:musen2s2}, then subtract it to produce a ``pure" H$\alpha$ image.  {\bf Bottom-center:} H$\beta$ generally traces H$\alpha$, but is negligible in the dust lane (NE from the nucleus) which runs parallel to the brightest line structure.  {\bf Bottom-right:} Two-color image of [\ion{S}{2}] (green) and H$\alpha$ (magenta) illustrating the relative lack of Balmer emission in the SW filamentary loop.  [\ion{S}{2}] which also traces bright H$\alpha$ appears white.  Additional [\ion{S}{2}]-bright filamentary structure is visible at large angles from the bicone, with examples of smaller structure indicated by yellow circles and a the edges of a possible large NE loop indicated by yellow dotted lines.} 
\label{fig:linepics}
\end{figure*}

\section{Results}

The continuum-subtracted images containing forbidden line emission ([\ion{O}{3}],  [\ion{S}{2}], H$\alpha$+[\ion{N}{2}]) are shown in Figure \ref{fig:linepics} (top).  Figure \ref{fig:linepics} (bottom) shows continuum-subtracted H$\beta$ and clean H$\alpha$ (with contaminating [\ion{N}{2}] modeled and subtracted), as well as a two-color comparison between [\ion{S}{2}] and H$\alpha$.  The WFPC2 image in Fig. \ref{fig:linepics} ([\ion{O}{3}]) is sampled at $0\farcs09$, and all others are at $0\farcs046$.

\subsection{An Ionized Forbidden Loop}
\label{sec:loop}

Fig. \ref{fig:linepics} shows that to first order, narrow line emission from H$\alpha$, H$\beta$ and [\ion{S}{2}] track the [\ion{O}{3}] emission \citep[previously described by][]{Schmitt03}.  A bright X-shaped ridge of line emission extends SE-NW, parallel to and south of a dust lane.  Several additional details are present in the new WFC3 images.  The most prominent new feature is a loop (possibly a bubble seen in projection) that extends SW from the nucleus, with minor axis $\lesssim1\farcs0$ and major axis $\lesssim3\farcs0$ (perpendicular to the main line emission structure).  This loop is present in [\ion{S}{2}] and blended H$\alpha$+[\ion{N}{2}] (Fig. \ref{fig:linepics}, top), but appears to be predominantly a low-ionization forbidden feature.  It is not present in [\ion{O}{3}] (Fig. \ref{fig:linepics}, top) or H$\beta$ (Fig. \ref{fig:linepics}, bottom), and it is not present in H$\alpha$ (Fig. \ref{fig:linepics}, bottom), implying the loop's contribution to the blended image is predominantly via [\ion{N}{2}].  The contrast between [\ion{S}{2}] and H$\alpha$ becomes obvious via direct two-color comparison (Fig. \ref{fig:linepics}, bottom right).  Integrating over the whole loop to improve significance, we find [\ion{O}{3}]/H$\beta$\,$\sim3.0$ and [\ion{S}{2}]/H$\alpha$\,$\sim1.3$ for typical flux and all values of [\ion{N}{2}]/[\ion{S}{2}].

\begin{figure}
\centering
\vspace{0.15in}
\noindent
\begin{overpic}[width=0.9\linewidth]{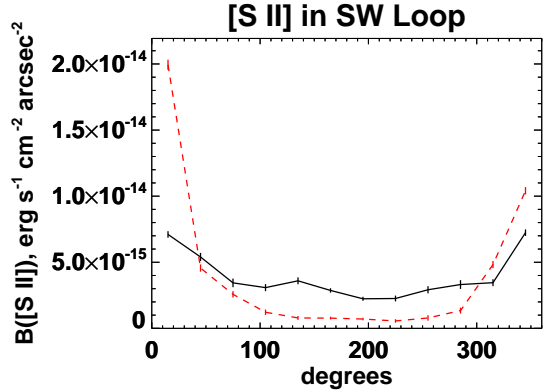}
\end{overpic}\\
\caption{Background-subtracted azimuthal profile of the loop (black, solid) and a slightly larger concentric elliptical annulus (red, dashed) immediately SW of the nucleus, as seen in [\ion{S}{2}] (Fig. \ref{fig:linepics}, top).  The peaks at $0\degr$ and $360\degr$ reflect planar material within the bicone.  Typical significance is $\sim10\sigma$ for each bin, so the loop is closed on $\simless30\degr$ scales.
}
\label{fig:loopaz}
\end{figure}

To determine if the loop is indeed a true loop (i.e. closed at all angles), we extract an azimuthal profile from the [\ion{S}{2}] images, colored black in Fig. \ref{fig:loopaz}.  Background is taken from the low-intensity ellipse at the center of the loop, and the profile itself consists of $30\degr$ segments of an elliptical annulus concentric with the loop ([$1\farcs44$,$0\farcs54$] semi-major axis [outer,inner] 18\degr\ East of North, and [$0\farcs72$,$0\farcs27$] semi-minor axis [outer, inner]).  [\ion{S}{2}] is detected from the loop at $\simgreat10\sigma$ for all bins, indicating that the loop is closed on scales approaching the limit of {\it HST} resolution.  For comparison, we also profile a slightly larger concentric elliptical annulus which does not contain emission from the loop.  This annulus, red in Fig. \ref{fig:loopaz}, has [$2\farcs20$,$1\farcs70$] semi-major axis [outer,inner], and [$1.10\farcs72$,$0\farcs85$] semi-minor axis [outer, inner]).  Regions of this second larger annulus are brighter than the loop due to the inclusion of planar material.

The relative intensity of the loop in [\ion{S}{2}] and H$\alpha$ depends strongly upon the assumed [\ion{N}{2}]/[\ion{S}{2}] ratio used for [\ion{N}{2}] subtraction, which may be subject to systematic errors in continuum subtraction and flux scaling.  Varying the assumed [\ion{N}{2}]/[\ion{S}{2}] ratio over the range of values shown from MUSE (Fig. \ref{fig:musen2s2}) increases the H$\alpha$ contribution for low values (the lowest value that we considered, [\ion{N}{2}]/[\ion{S}{2}]$=1.35$, corresponding to $\sim1000\,\kms$ shocks, is representative for the loop region).  H$\alpha$ images generated using low values of [\ion{N}{2}]/[\ion{S}{2}]$=[1.35,1.58]$ do show hints of a loop in H$\alpha$, particularly towards the nucleus.  But [\ion{S}{2}] remains dominant, even under these more conservative assumptions, as is evident from the LINER structure associated with the loop in Fig. \ref{fig:bptlo}.

\begin{figure*}
\setlength{\fboxsep}{5pt}
\centering
\noindent
\fbox{
\begin{overpic}[width=0.268\linewidth,trim=0 0 1.1cm 0,clip]{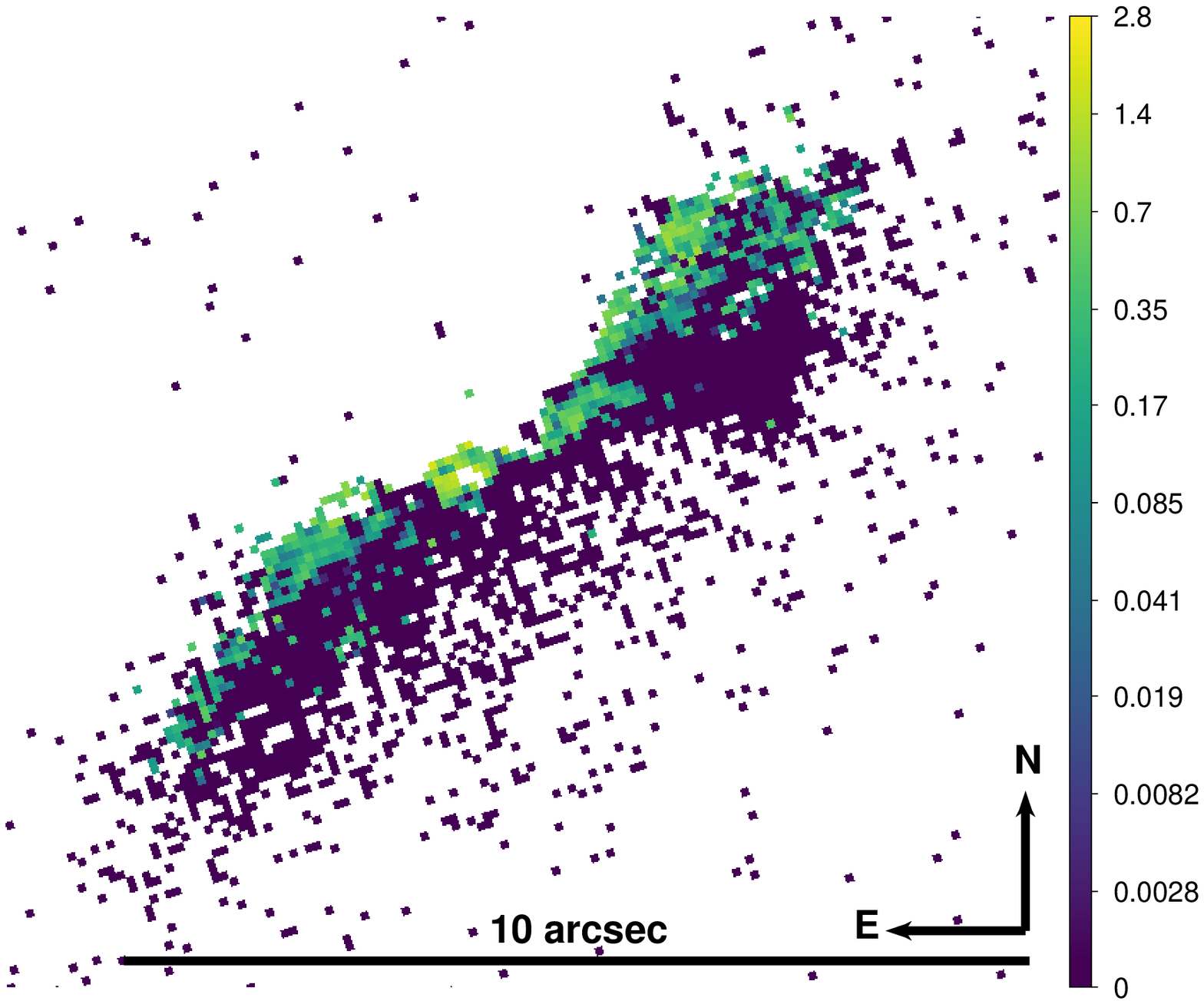}
	\put(3,73){\color{white}\rule{2.5cm}{0.5cm}}
	\put(3,77){{\parbox{2cm}{%
			\begin{equation}
			\rm{E(B-V),}\ 3\sigma \nonumber
			\end{equation}
			}}}
\end{overpic}\hspace{0.01\textwidth}}%
\fbox{
\begin{overpic}[width=0.268\linewidth,trim=0 0 1.1cm 0,clip]{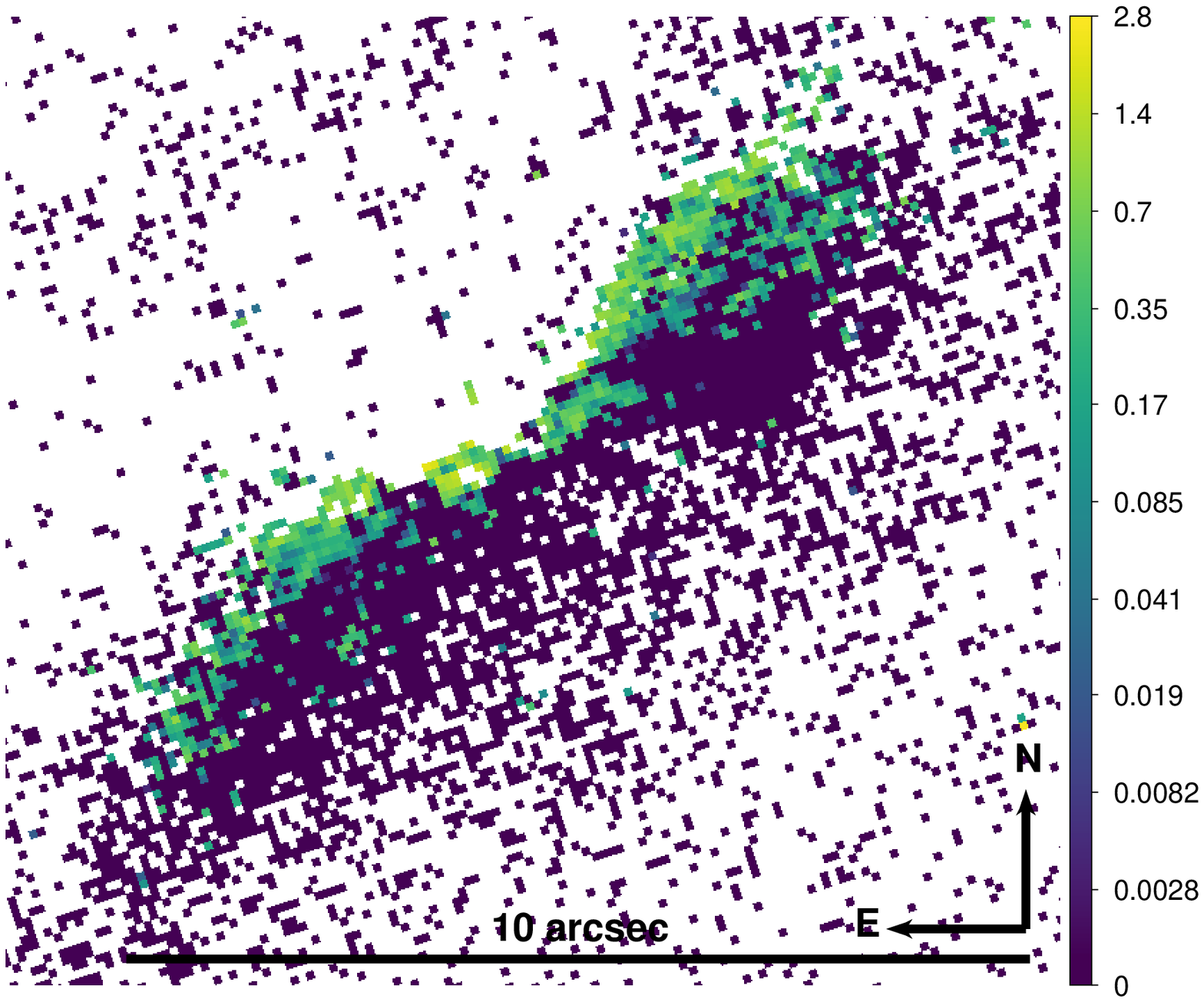}
	\put(3,73){\color{white}\rule{3.3cm}{0.5cm}}
	\put(3,77){{\parbox{2cm}{%
			\begin{equation}
			\rm{E(B-V),\,adaptive} \nonumber
			\end{equation}
			}}}
\end{overpic}\hspace{0.01\textwidth}}%
\fbox{
\begin{overpic}[width=0.305\linewidth]{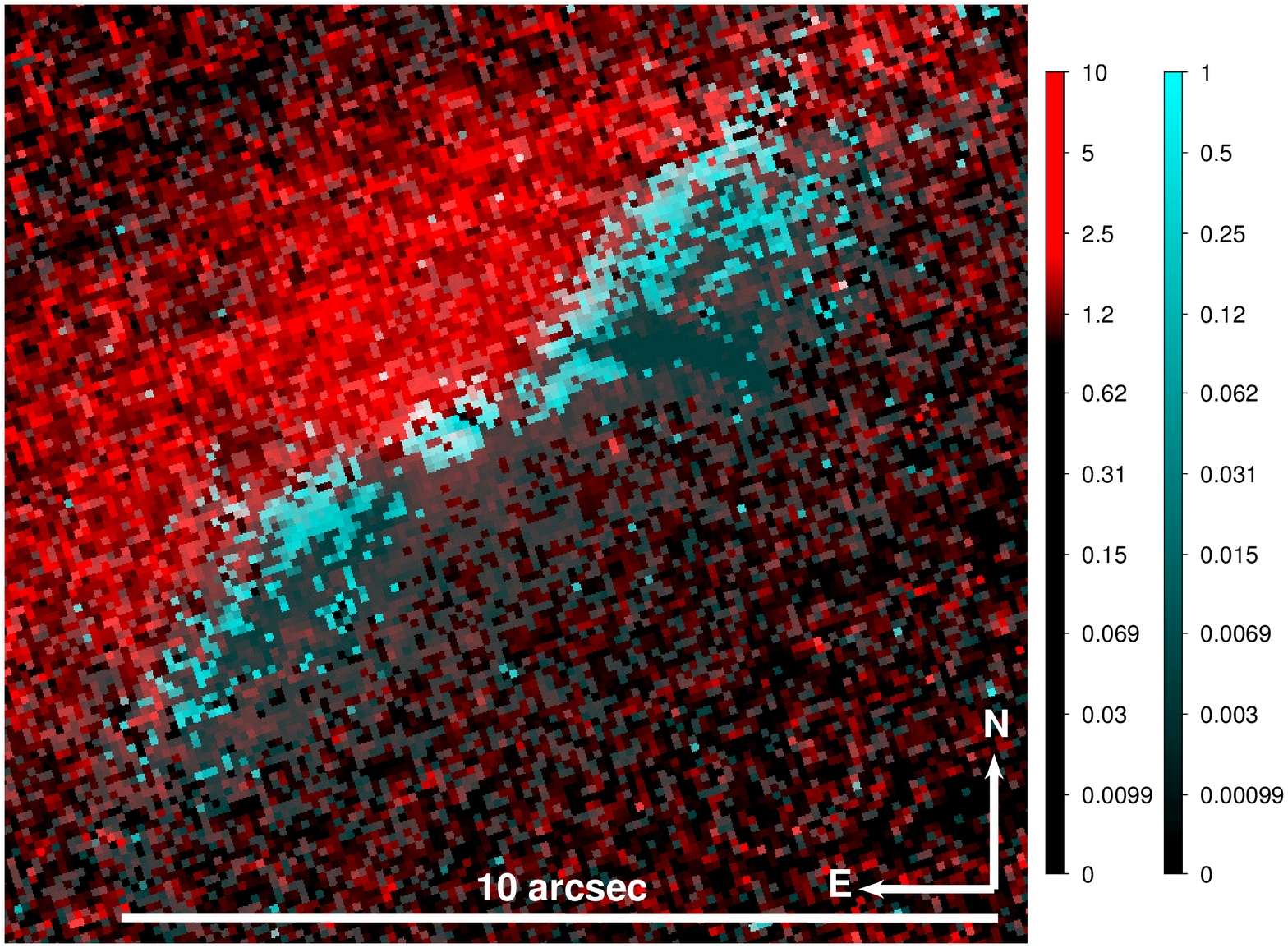}
	\put(1,65){\color{white}\rule{3.1cm}{0.5cm}}
	\put(2,69){{\parbox{2cm}{%
			\begin{equation}
			\rm{E(B-V)\,vs.\,color}  \nonumber
			\end{equation}
			}}}
\end{overpic}}\\
\caption{Extinction maps of the IC 5063 nucleus.  {\bf Left:} E(B-V) map (magnitudes) derived from H$\alpha$, H$\beta$ (Fig. \ref{fig:linepics}, bottom).  Pixels with H$\alpha$ or H$\beta$ below 3$\sigma$ are masked as white.  {\bf Center:} The same, but pixels are adaptively smoothed to retain 3$\sigma$ features.  Pixels with H$\alpha$ or H$\beta$ below 1$\sigma$ are masked as white. 
{\bf Right:} Two-color extinction map comparing continuum color (F763M/F547M; red) and E(B-V) (green).  The most extincted regions (brightest red) are heavily extincted by the nuclear dust lane, so significant H$\alpha$ or H$\beta$ cannot be measured.}
\label{fig:EBV}
\end{figure*}

\subsection{Extinction Effects}

The limits of the narrow line images are strongly affected by the dust lane that crosses the nucleus NW-SE, parallel to the NLR along its NE edge.  As in \cite{Kreckel13}, we convert H$\alpha$ and H$\beta$ into extinction (Fig. \ref{fig:EBV}).  Extinction is nearly negligible across most of the NLR, but rises sharply to E(B-V)$\simgreat1.0$ approaching the dust lane.  NE of that, 
H$\alpha$ and H$\beta$ are too faint to be used.  For comparison, we map the F763M/F547M continuum ratio (comparable to V-I), which remains comparably high up to $\simless2\farcs4$ NE of the nucleus.

\begin{figure*}
\centering
\vspace{0.15in}
\noindent
\begin{overpic}[width=0.43\linewidth]{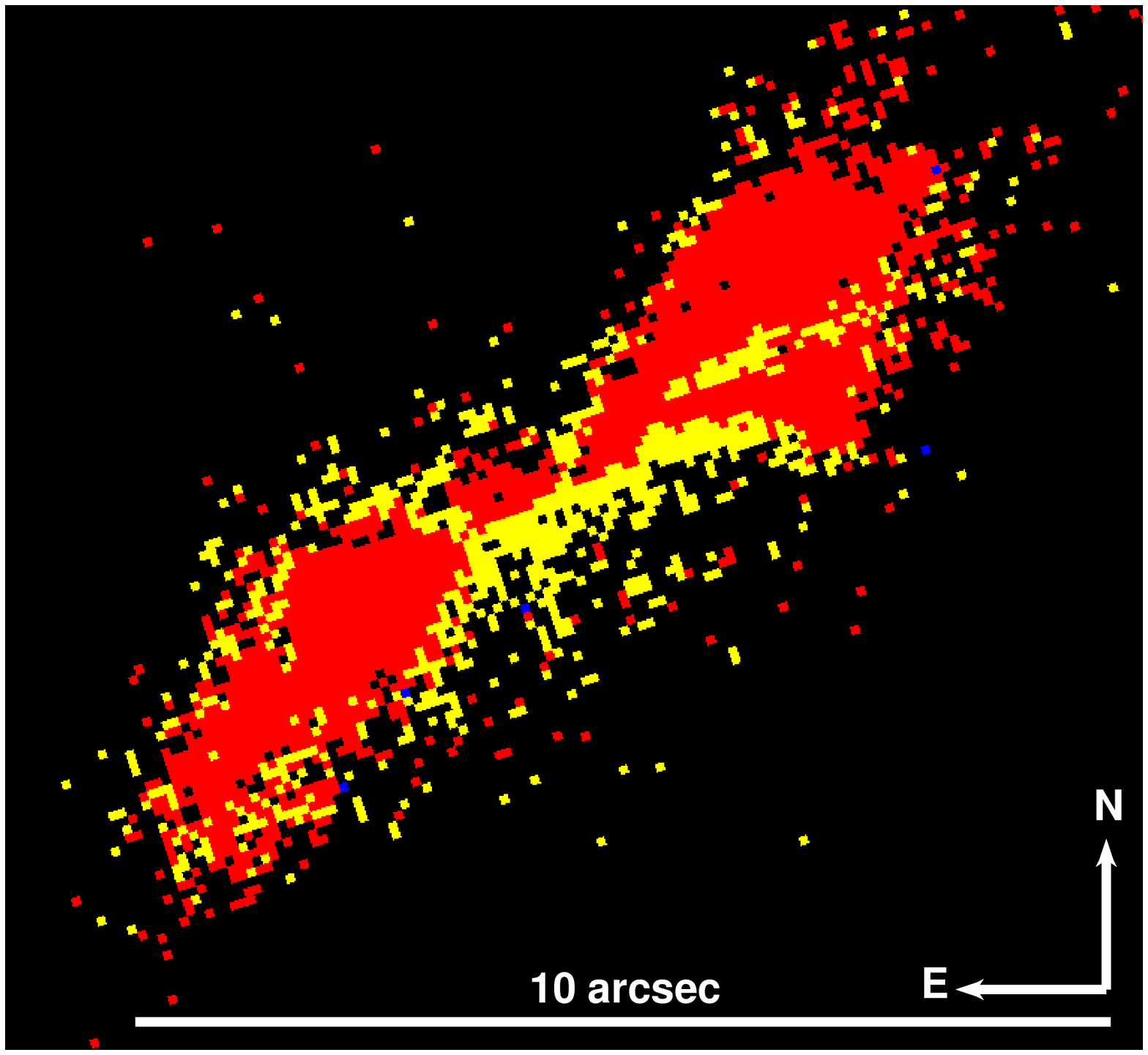}
	\put(45,31){\color{white}\circle{5}}
	\put(55,41){\color{white}\circle{5}}
	\put(59,48){\color{white}\circle{5}}
	\put(3,80){\color{white}\rule{5.5cm}{0.5cm}}
	\put(3,83){{\parbox{5.5cm}{%
			\begin{equation}
			{\rm BPT\ map,\ } 3\sigma \nonumber, [\rm{N\,II}/\rm{S\,II}]=1.81
			\end{equation}
			}}}
\end{overpic}\hspace{0.05\textwidth}%
\begin{overpic}[width=0.43\linewidth]{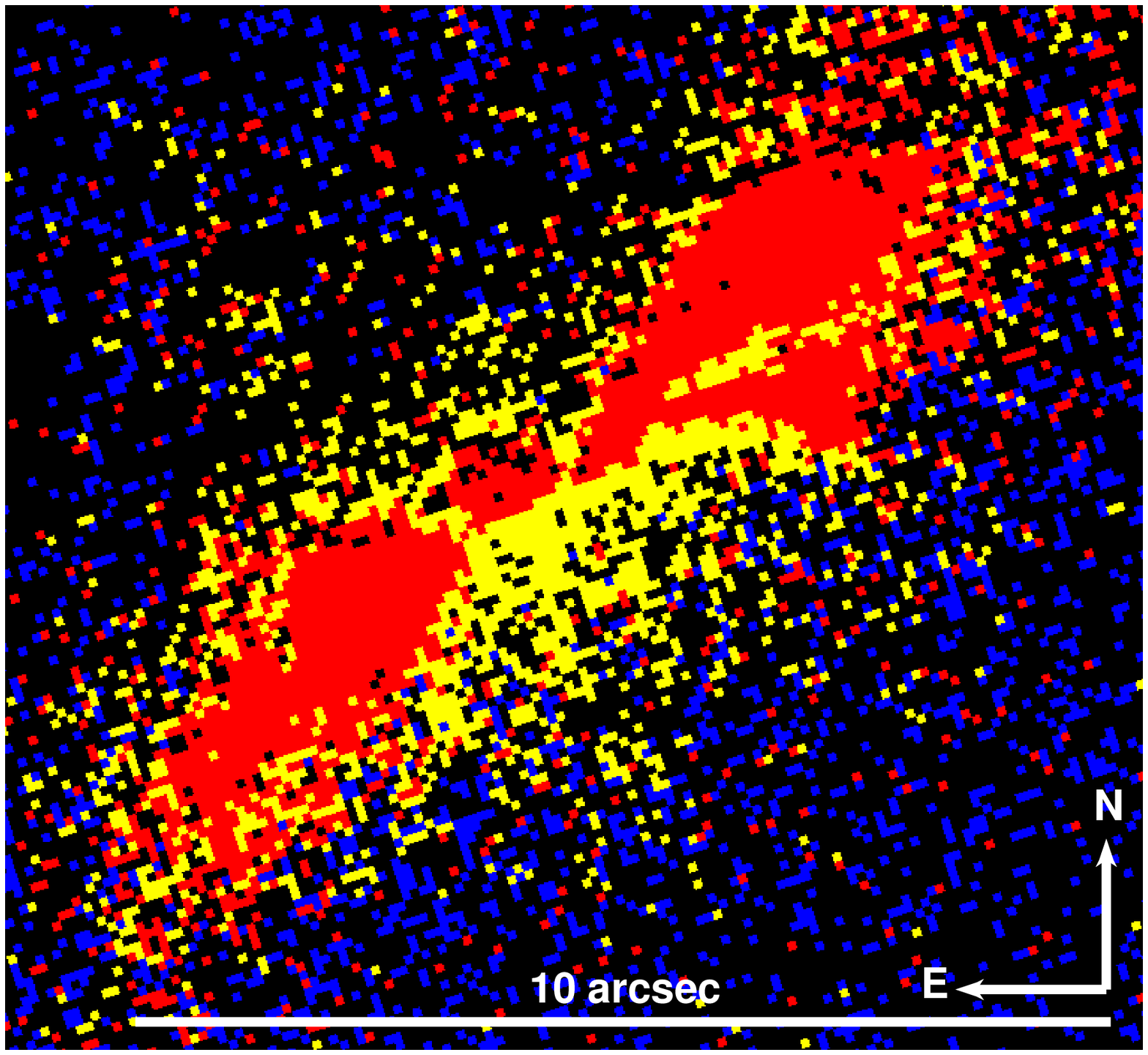}
	\put(23,64){\color{white}\circle{5}}
	\put(46,31){\color{white}\circle{5}}
	\put(3,80){\color{white}\rule{6.5cm}{0.5cm}}
	\put(3,82){{\parbox{6.5cm}{%
			BPT map, adaptive, [\rm{N\,II}/\rm{S\,II}]=1.81
			}}}
\end{overpic}\\
\vspace{0.1in}
\noindent
\begin{overpic}[width=0.47\linewidth,trim=0 0 0 0,clip]{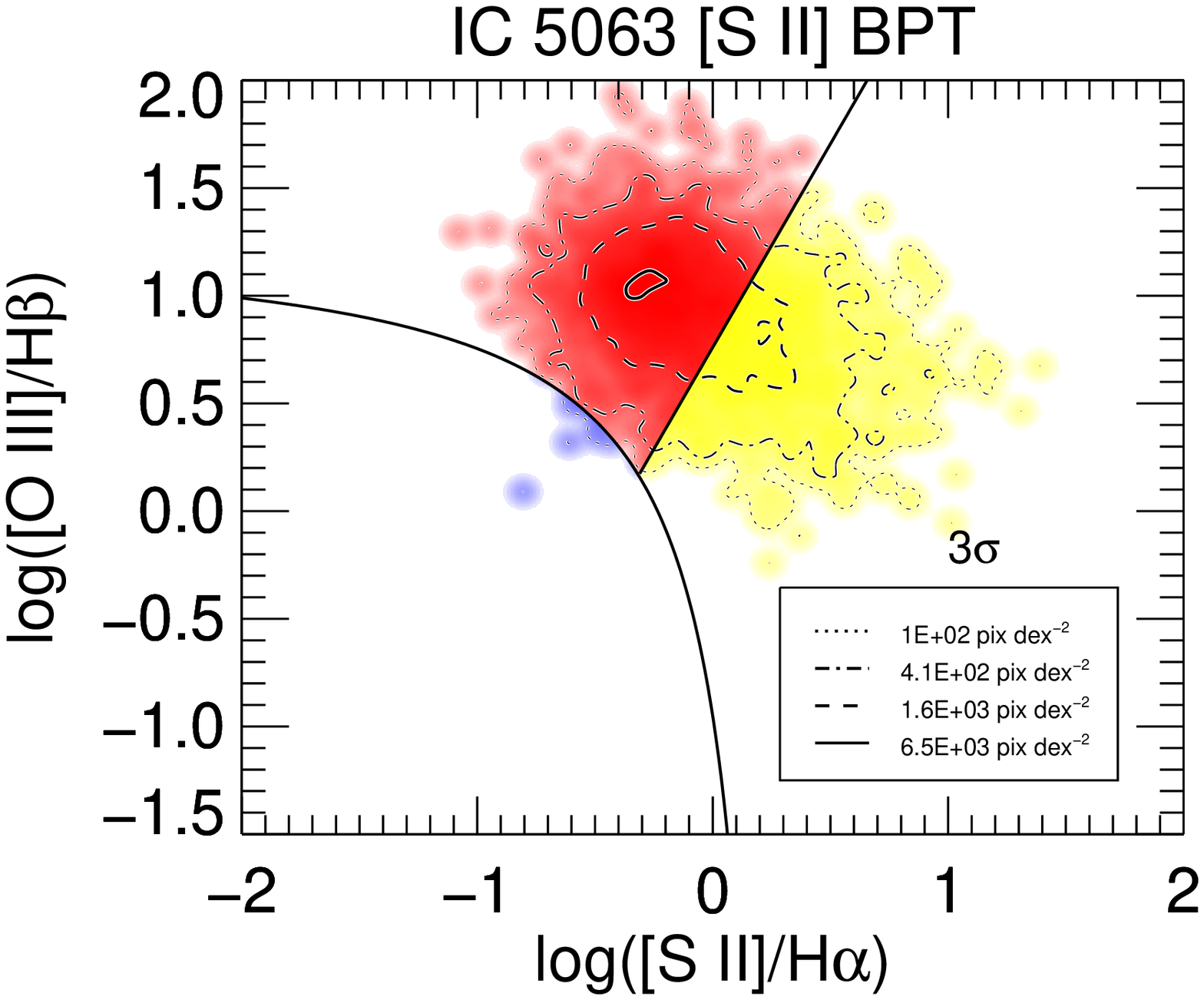}
\end{overpic}\hspace{0.02\textwidth}%
\begin{overpic}[width=0.47\linewidth,trim=0 0 0 0,clip]{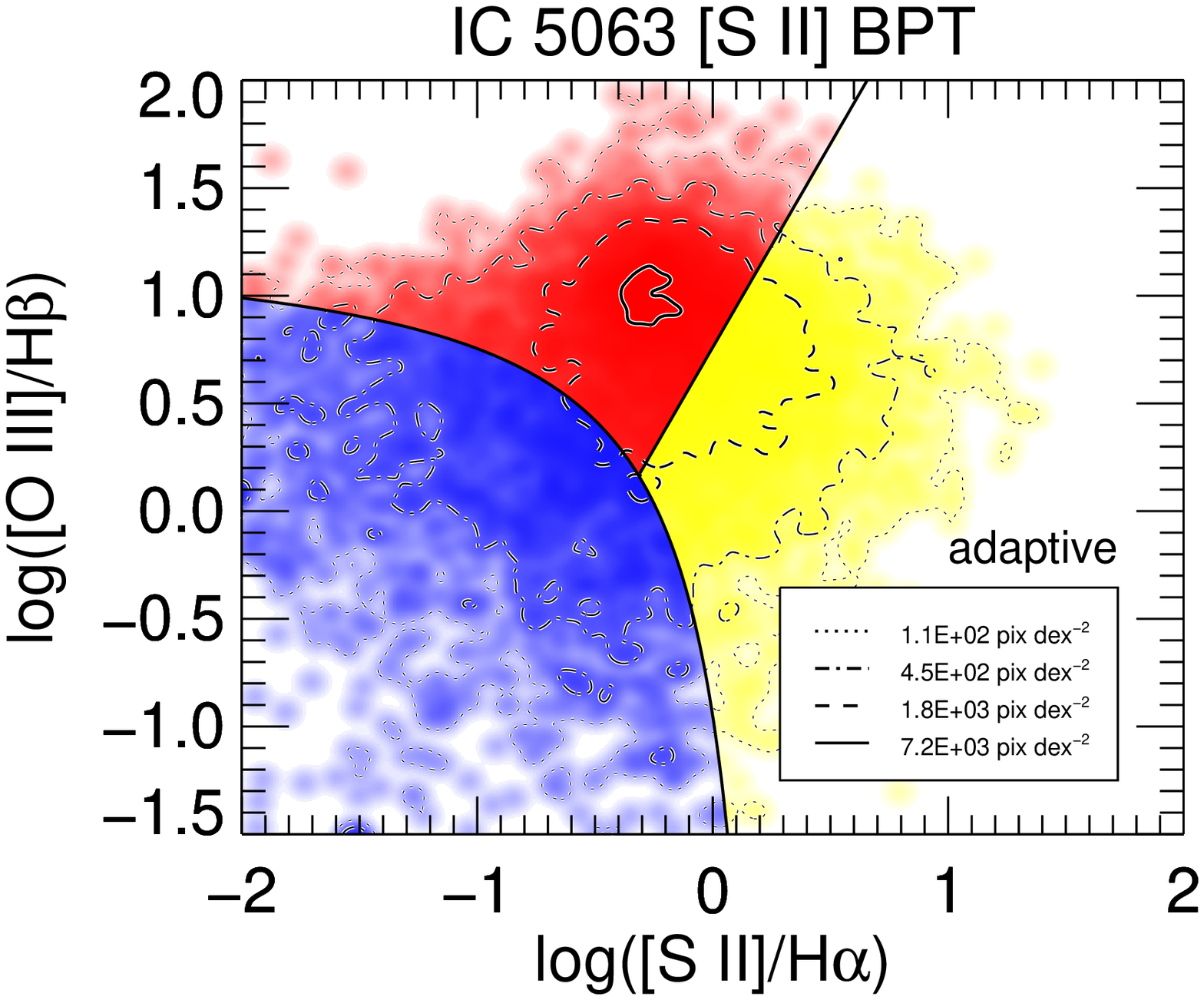}
\end{overpic}\\
\caption{{\bf Top:} Baldwin-Phillips-Terlevich (BPT; 1981) diagnostic maps of the IC 5063 nucleus, using continuum- and contamination-subtracted {\it HST} narrow line images from Fig. \ref{fig:linepics}.  Each HST pixel is categorized based upon those line ratios.  Red is Seyfert-like, yellow is LINER-like, and blue indicates ratios typical of star-forming (SF) regions.  Black pixels have been masked due to at least one non-significant line.  A few larger ($\sim2-4$ pixel span, $\sim40-80$\,pc) examples of highly ionized clumps in the cross-cone are circled in white.  {\bf Top-left:} non-significant line values have been masked below 3$\sigma$. {\bf Top-right:} Line maps have been adaptively smoothed for pixels between 1-3$\sigma$.  Pixels with a line below 1$\sigma$ significance are masked.  As in \cite{Maksym16}, Seyfert-like emission is primarily confined to a narrow angle within the bicone.  The NE cross-cone is obscured by the dust lane, so line measurements depend upon a reddening correction.  The SW cross-cone contains patchy LINER-like emission, as well as SF-like emission visible in the adaptive image.  {\bf Bottom:} BPT diagrams of the inner 10\arcsec\ of IC 5063.  Each $0\farcs09\times0\farcs09$ pixel is a single data point.  In order to maintain clarity for both the most and least densely populated parts of the diagram, the data points are binned to a 2-D histogram and smoothed by Gaussian.  Classification criteria established by \citep{Kewley06} are indicated as solid lines, and colors are as in the top.  Contours indicate smoothed phase space density along the line ratio axes, per square dex.  {\bf Bottom-left:} Data use a 3$\sigma$ cut, as in the top-left panel.  The bright pixels which are retained by the cut form a continuum that runs perpendicular to the LINER-Seyfert divide, and contains little SF.  {\bf Bottom-right:} Data are adaptively smoothed, as in top-right panel. Comparison with the top-right shows that most of the SF and many of the LINER-like regions are excluded with a 3$\sigma$ cut, but retained with adaptive smoothing.
}
\label{fig:bpt}
\end{figure*}

\begin{figure*}
\centering
\vspace{0.15in}
\noindent
\begin{overpic}[width=0.43\linewidth]{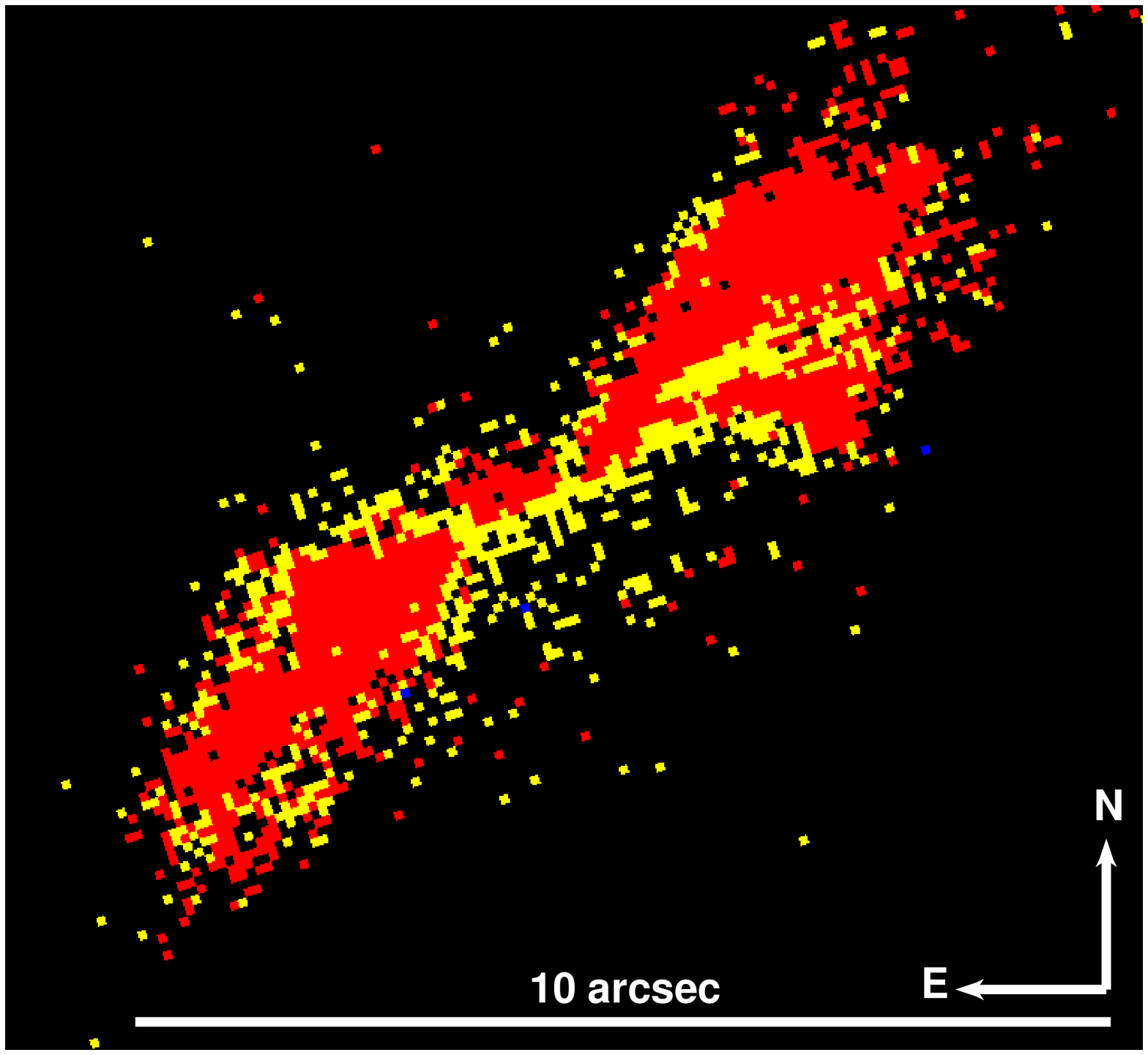}
	\put(3,80){\color{white}\rule{5.5cm}{0.5cm}}
	\put(3,83){{\parbox{5.5cm}{%
			\begin{equation}
			{\rm BPT\ map,\ } 3\sigma \nonumber, [\rm{N\,II}/\rm{S\,II}]=2.05
			\end{equation}
			}}}
\end{overpic}\hspace{0.05\textwidth}%
\begin{overpic}[width=0.43\linewidth]{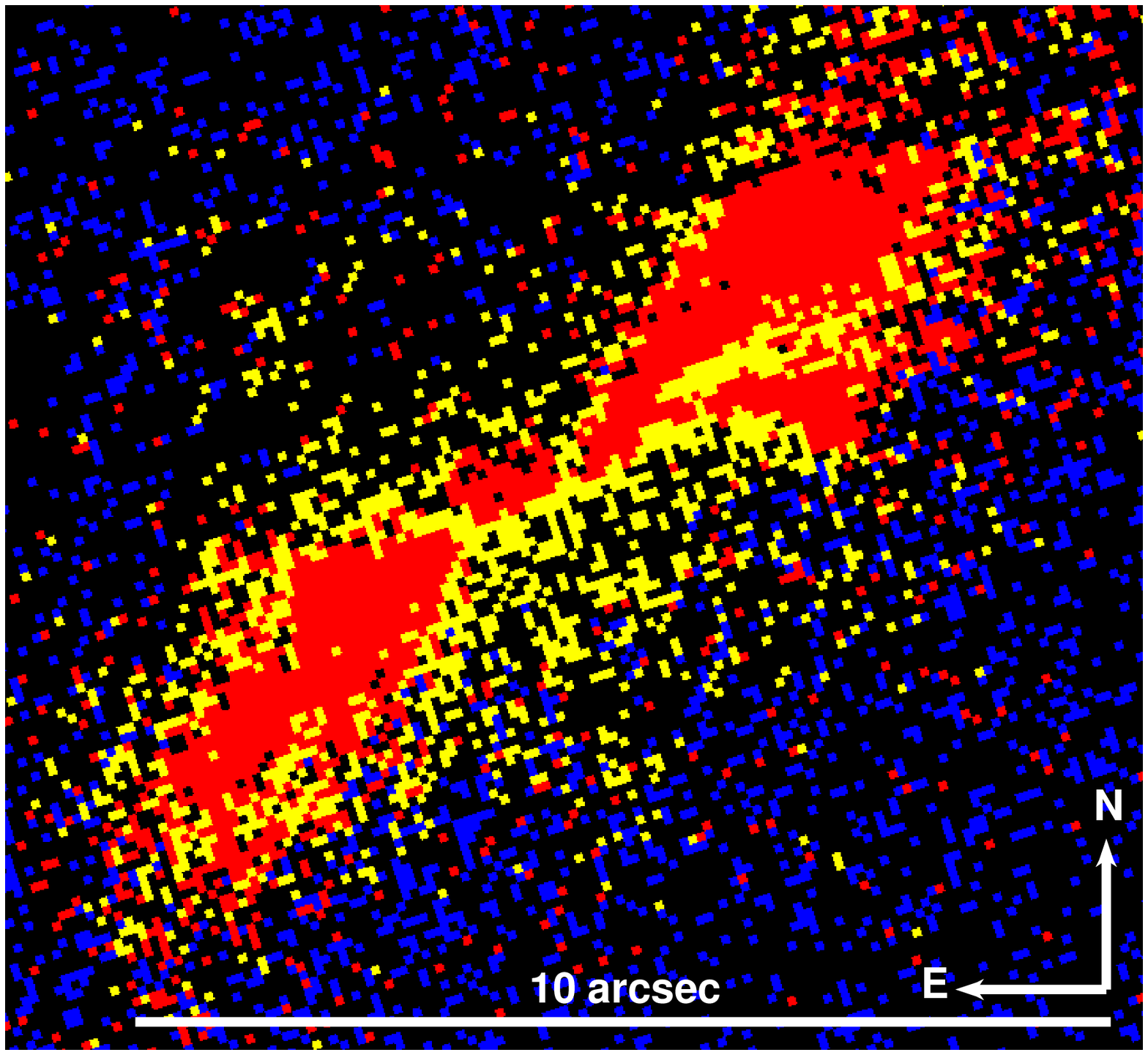}
	\put(3,80){\color{white}\rule{6.5cm}{0.5cm}}
	\put(3,82){{\parbox{6.5cm}{%
			BPT map, adaptive, [\rm{N\,II}/\rm{S\,II}]=2.05
			}}}
\end{overpic}\\
\vspace{0.1in}
\noindent
\begin{overpic}[width=0.47\linewidth,trim=0 0 0 0,clip]{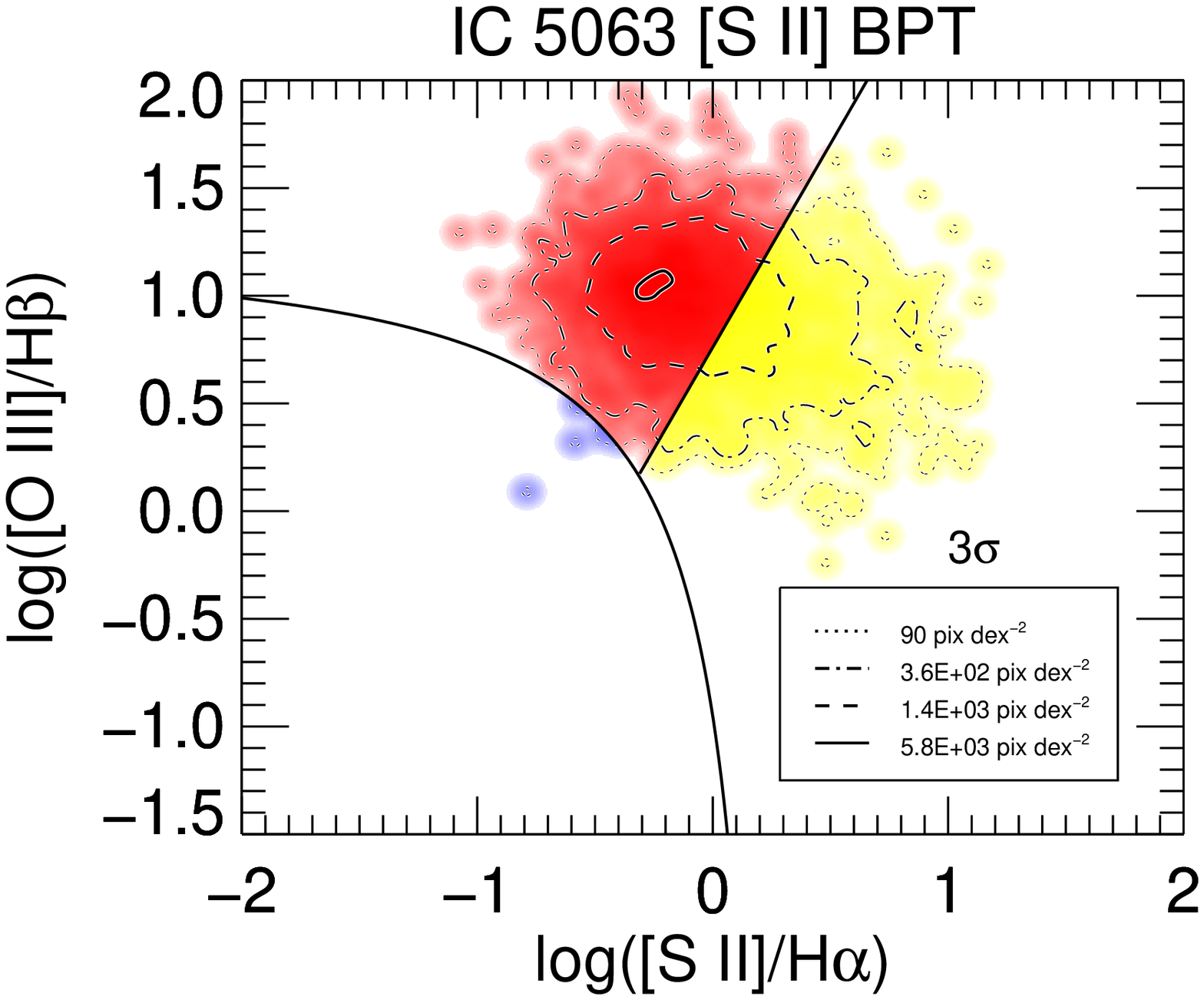}
\end{overpic}\hspace{0.02\textwidth}%
\begin{overpic}[width=0.47\linewidth,trim=0 0 0 0,clip]{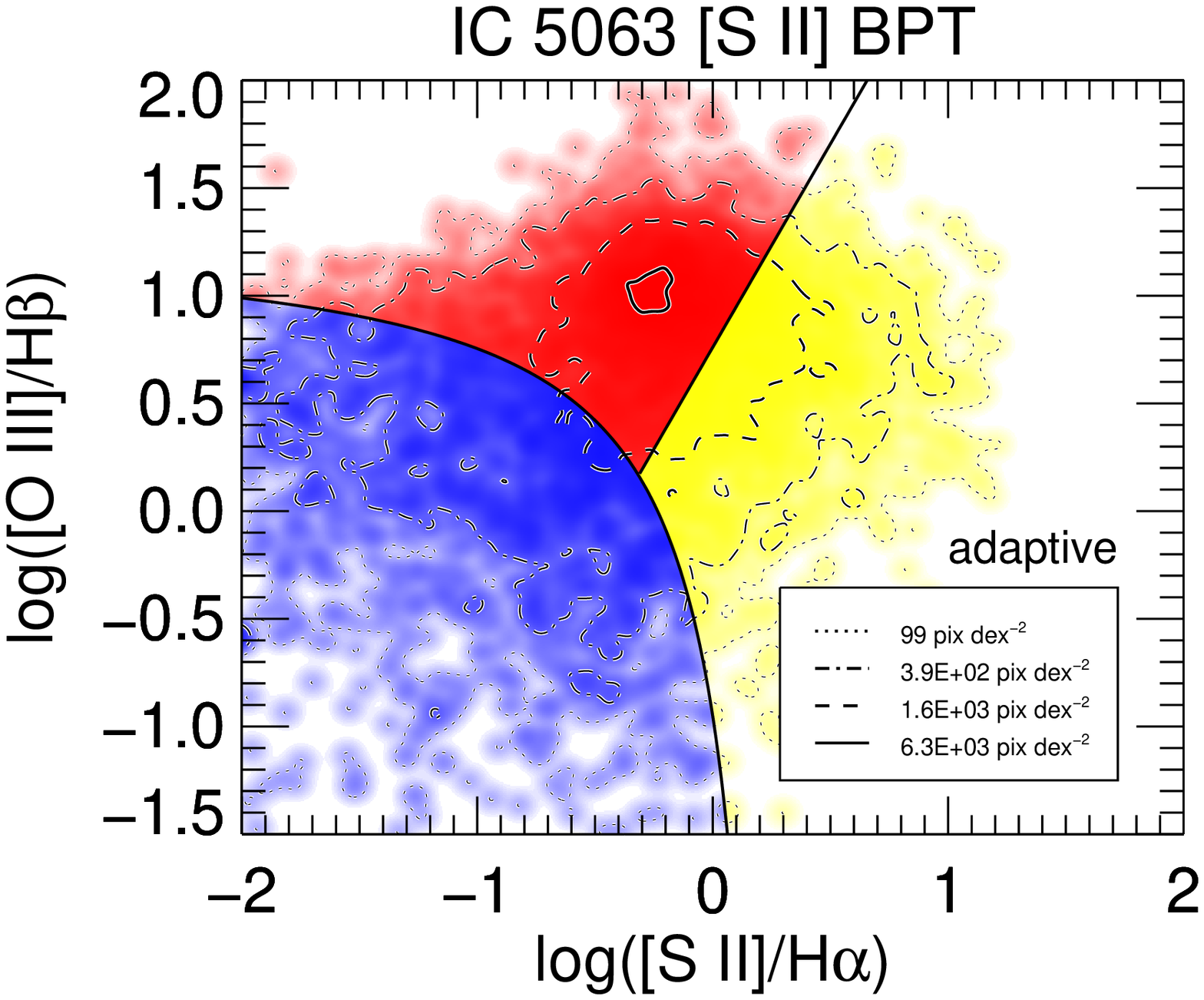}
\end{overpic}\\
\caption{As for Fig. \ref{fig:bpt}, but for [\ion{N}{2}]/[\ion{S}{2}]=2.05, comparable to the highest physically expected ratio values for either photoionization or shocks, and for high values in Fig. \ref{fig:musen2s2}.  Differences from Fig. \ref{fig:bpt} are minor.
}
\label{fig:bpthi}
\end{figure*}

\begin{figure*}
\centering
\vspace{0.15in}
\noindent
\begin{overpic}[width=0.43\linewidth]{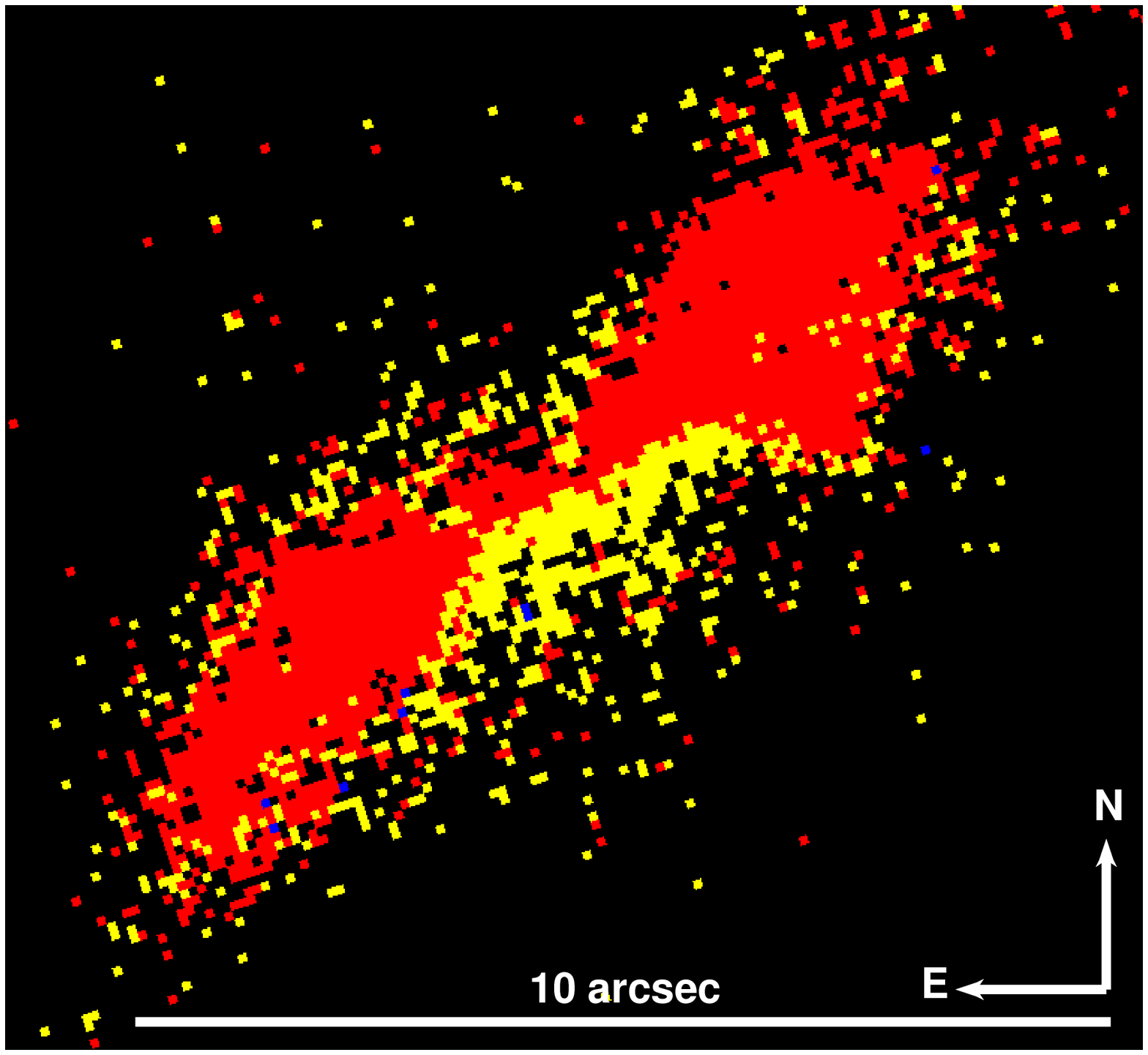}
	\put(3,80){\color{white}\rule{5.5cm}{0.5cm}}
	\put(3,83){{\parbox{5.5cm}{%
			\begin{equation}
			{\rm BPT\ map,\ } 3\sigma \nonumber, [\rm{N\,II}/\rm{S\,II}]=1.35
			\end{equation}
			}}}
\end{overpic}\hspace{0.05\textwidth}%
\begin{overpic}[width=0.43\linewidth]{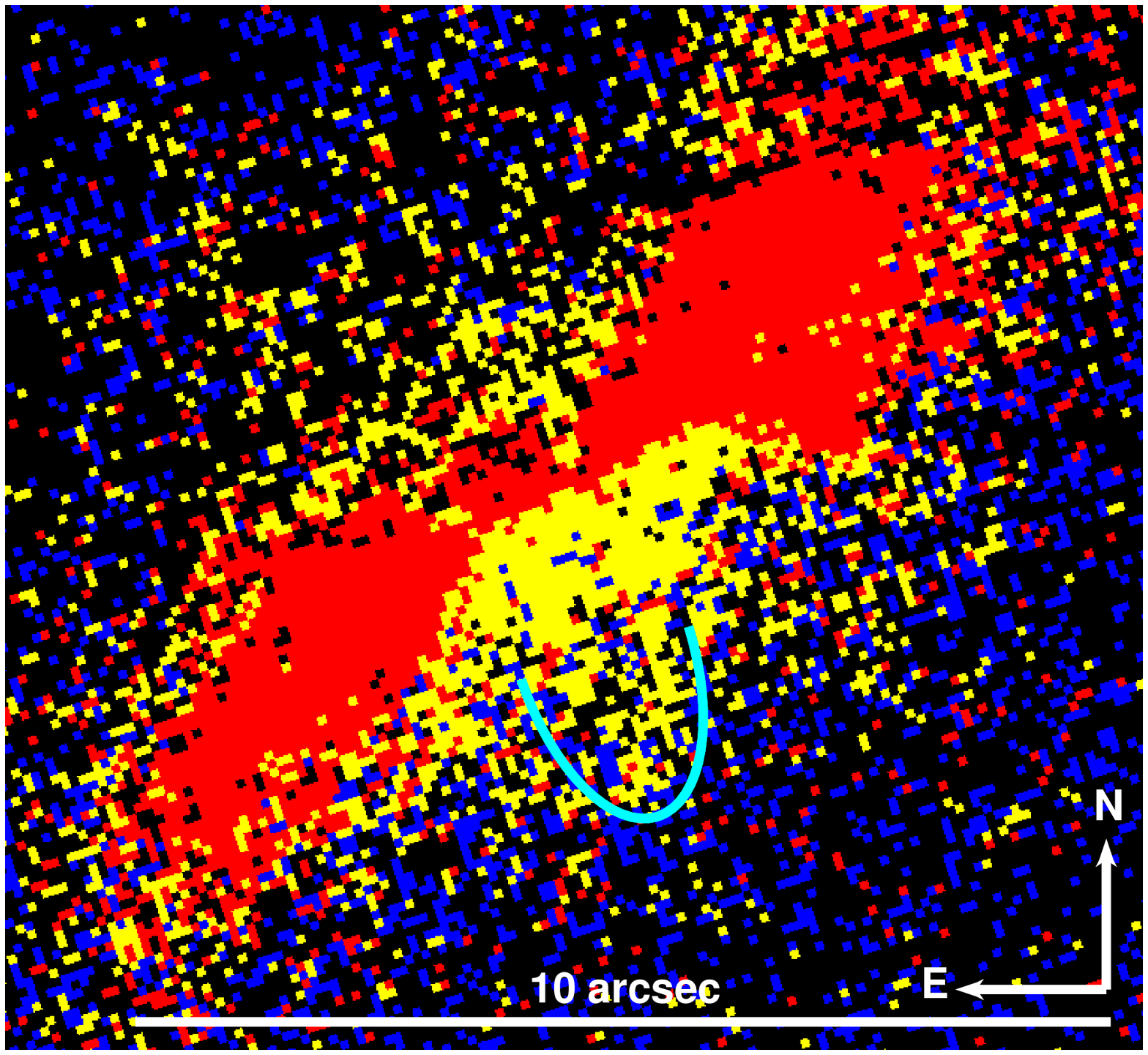}
	\put(3,80){\color{white}\rule{6.5cm}{0.5cm}}
	\put(3,82){{\parbox{6.5cm}{%
			BPT map, adaptive, [\rm{N\,II}/\rm{S\,II}]=1.35
			}}}
\end{overpic}\\
\vspace{0.1in}
\noindent
\begin{overpic}[width=0.47\linewidth,trim=0 0 0 0,clip]{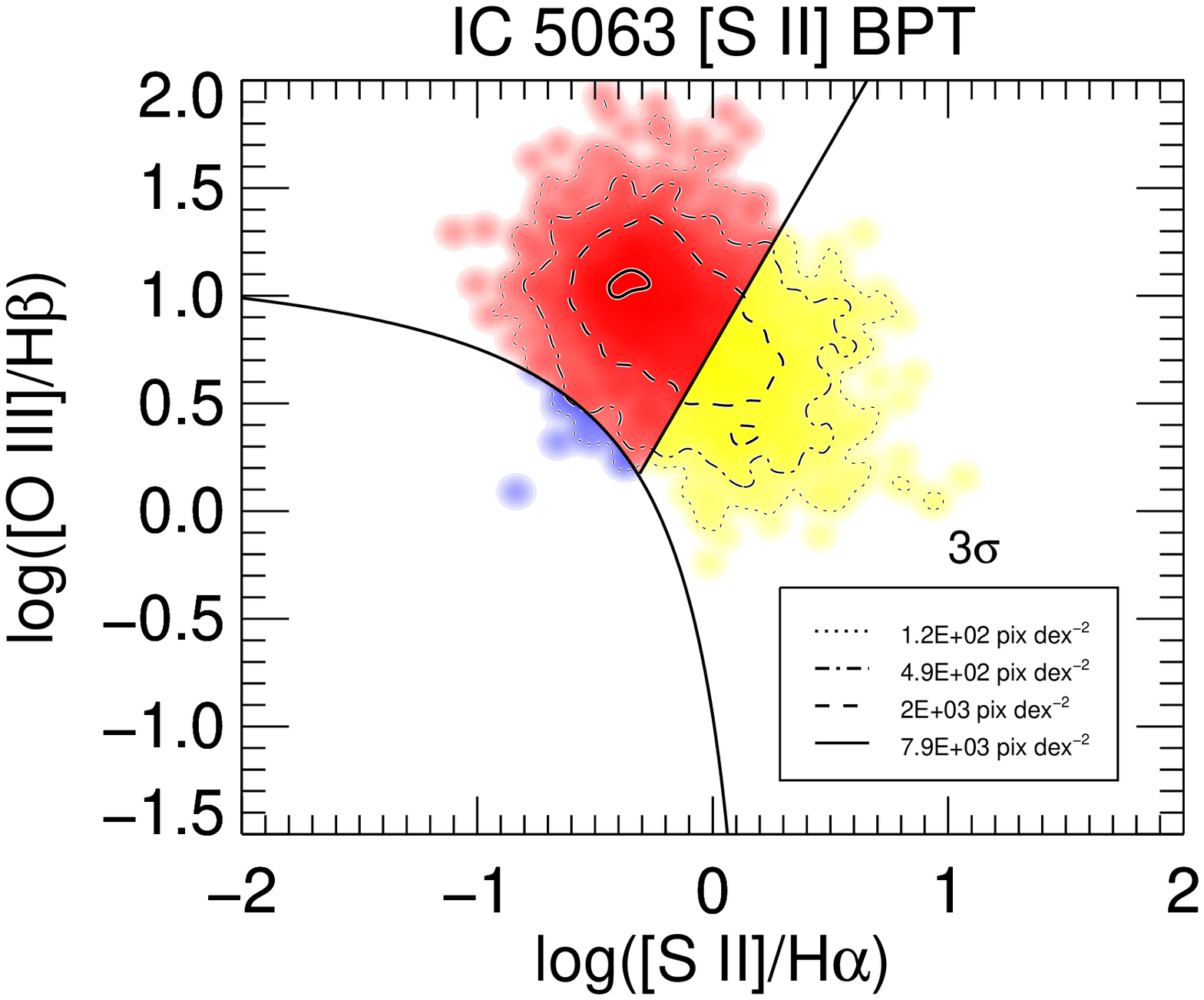}
\end{overpic}\hspace{0.02\textwidth}%
\begin{overpic}[width=0.47\linewidth,trim=0 0 0 0,clip]{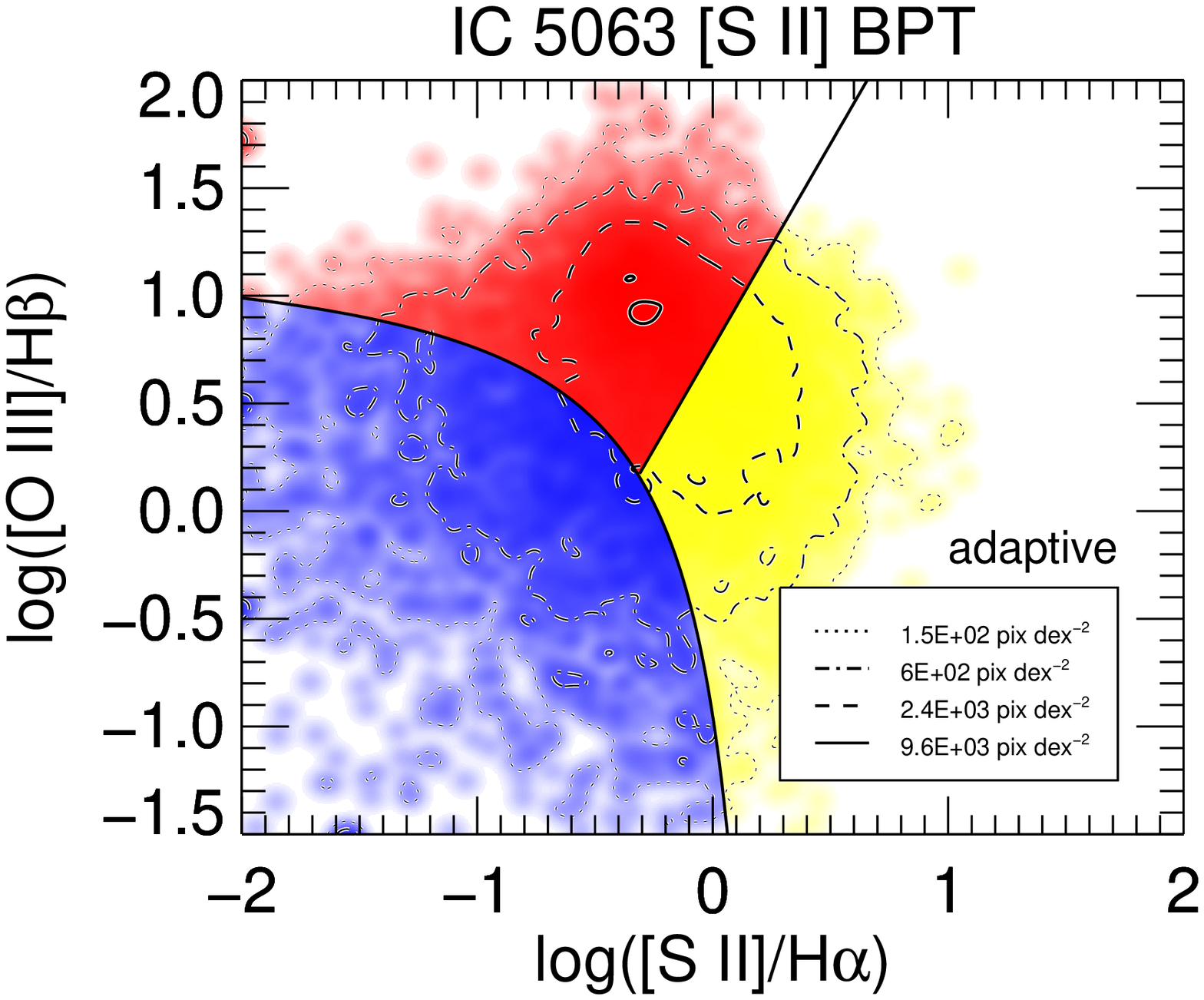}
\end{overpic}\\
\caption{As for Fig. \ref{fig:bpt}, but for [\ion{N}{2}]/[\ion{S}{2}]=1.35, comparable to the lowest physically expected ratio values (for fast shocks), and for low observed values in Fig. \ref{fig:musen2s2}.  In the adaptively smoothed map (unlike in Fig. \ref{fig:bpt}), there is no LINER/SF ``notch" where the NLR bifurcates in the NW.  The the SW ``loop" seen primarily in [\ion{S}{2}] (Fig. \ref{fig:linepics}; marked here with a green half-ellipse) becomes prominently LINER-like under adaptive smoothing.
}
\label{fig:bptlo}
\end{figure*}

\subsection{NLR Extent}

The size of the NLR is connected to the AGN's ability to excite the surrounding gas, and hence to its power.  In order to compare against previous studies studying the observed relationship between total [\ion{O}{3}] emission and NLR extent \citep[e.g.][and subsequent studies]{Schmitt03,Bennert06,Greene11},  we follow the prescription of \cite{SB18} to measure the extent $R_{maj}$ as half the maximum distance spanning the $3\sigma$ flux isocontour.  We then sum the flux $F_{[\rm{O\,III}]}$ within this contour over the total bandwidth (assuming 55\AA\ from the FR533N PHOTBW header key).    

We find $R_{maj}=1.4\,\rm{kpc}$ and $F_{[\rm{O\,III}]}=6.88\times10^{-13}\,$\ecmss\ for a $3\sigma$ contour of $1.43\times10^{-15}\,\icgsbri$.  This implies $L_{[\rm{O\,III}]}=1.89\times10^{41}\,\es$. This is comparable to the values found by \citep{Schmitt03} which are commonly used in subsequent studies of $L_{[\rm{O\,III}]}-R_{maj}$.  It is somewhat above the best-fit values from \citep{SB18} (log$[R_{maj}/\rm{pc}]$=3.15, vs. regression values of $2.71\pm1.7$ for Type 2 AGN and $2.93\pm1.6$ for all AGN).  We note, however, that the limiting surface brightness sensitivity can produce wildly varying results.  For comparison, [\ion{O}{3}] in the bicone spans the entire MUSE datacube at $3\sigma$ and may extend beyond it ($R_{maj}=8.6$\,kpc, 37\arcsec).

\subsection{Seyfert-like and LINER-like Regions}

The BPT maps generated from [\ion{O}{3}]/H$\beta$ and [\ion{S}{2}]/H$\alpha$ ratio maps are depicted in Figs. \ref{fig:bpt}, \ref{fig:bpthi} and \ref{fig:bptlo} (top) for H$\alpha$ maps derived using the median (1.82), high (2.05), and low (1.35) values of [\ion{N}{2}]/[\ion{S}{2}]. Varying  [\ion{N}{2}]/[\ion{S}{2}] has little effect in the 3$\sigma$ maps, but can be important for extended structure in the adaptively smoothed maps where much of the emission coincides with low [\ion{N}{2}]/[\ion{S}{2}] values in Fig. \ref{fig:bptlo}.  Very low [\ion{N}{2}]/[\ion{S}{2}] values tend to increase H$\alpha$, which obviously shifts bounds of the most extreme [\ion{S}{2}]/H$\alpha$ values to the left in the BPT diagrams.  Typical $1\sigma$ statistical uncertainties are $\simless0.2$ dex.

We see that the X-shaped ridge that dominates the nuclear line emission has predominantly Seyfert-like line ratios.  But at single-pixel resolution we identify several areas of LINER-like emission.  These LINER-like regions include a patchy strip along the southern edge of the ridge (opposite the dust lane), and along the inside of the NW notch where the `X' structure bifurcates.  This ``notch" LINER structure disappears only for [\ion{N}{2}]/[\ion{S}{2}]=1.35 (Fig. \ref{fig:bptlo}), which is not applicable to this region. Although much of the emission outside the bicone is too faint to benefit from adaptive smoothing, the smoothing does reveal additional LINER structure beyond this patchy strip, and SF emission which dominates with increasing radius.  This LINER emission becomes more evident in Fig. \ref{fig:bptlo} where with lower [\ion{N}{2}]/[\ion{S}{2}]=1.35, and which is a better match for this region in the MUSE ratio map (Fig. \ref{fig:musen2s2}).  This additional LINER structure is most obvious in the region containing the [\ion{S}{2}] loop.

We have plotted each {\it HST} pixel for regions within 10\arcsec\ of the nucleus in BPT diagrams (Figs. \ref{fig:bpt}, \ref{fig:bpthi} and \ref{fig:bptlo}, bottom), such that each point is located on [\ion{O}{3}]/H$\beta$ vs. [\ion{S}{2}]/H$\alpha$.   A comparison between the two cuts show that in all cases much of the SF is excluded by a $3\sigma$ cut, and retained with adaptive smoothing.  The brighter Seyfert-like and LINER-like pixels form a continuum that runs roughly perpendicular to the Seyfert-LINER divide described by \cite{Kewley06}, and seen in \cite{Maksym16,Ma21}.

\section{Discussion}


The most spectacular ionization cones are commonly inclined towards the plane \citep[e.g. NGC 1068 and Circinus.  See][]{VBH97}.  This configuration is well-established in IC 5063.  The 8 GHz radio continuum runs along the dust lane, with radio knots at the nucleus and two optically bright sites of bright line emission \citep{Morganti98,Kulkarni98}.  The physical association between the radio jet and line emission from optically thin molecular and ionized atomic gas suggests jet-ISM interactions, which are likely to produce shocks given the gas large velocities (e.g. $\sim600-1200\,\kms$ in NIR spectra, \citealt{Dasyra15}).  Bright [\ion{Fe}{2}] and H$_2$ emission in particular points to shocks at these sites \citep{Kulkarni98}.  Recent ALMA observations show that high excitation molecular gas is coincident with the radio jet and suggest that fast shocks caused by the jet are inflating a cocoon in the ISM and driving a lateral outflow \citep{Oosterloo17}.

The optical lines that we investigate have been studied in detail by \cite{Morganti07} via slit spectroscopy along the bicone axis and show complex structure with multiple kinematic components which point to strong interactions between the radio plasma and the ISM.  So although the spatial resolution of this study has major advantages ($0\farcs09$ pixels vs. $0\farcs8$ slit width), in practice it is not possible to avoid blending the multiple kinematic components, which have comparable intensity, cover a wide velocity range ($100\,\kms<\rm{FWHM}<1300\,\kms$).  Such components may also vary significantly in ionization, density and extinction ($4.6<$[\ion{O}{3}/H$\beta]<12.1$, $100\,\rm{cm}^{-3}<n<3000\,\rm{cm}^{-3}$ and $0.82<E(B-V)<1.70$ within individual extraction regions).  Such complexity is common in AGN outflows that drive molecular gas \citep[e.g.][]{Revalski18}.

\subsection{BPT Mapping and LINER-like Structures}

\begin{figure*}
\centering
\vspace{0.15in}
\noindent
\includegraphics[width=0.95\linewidth]{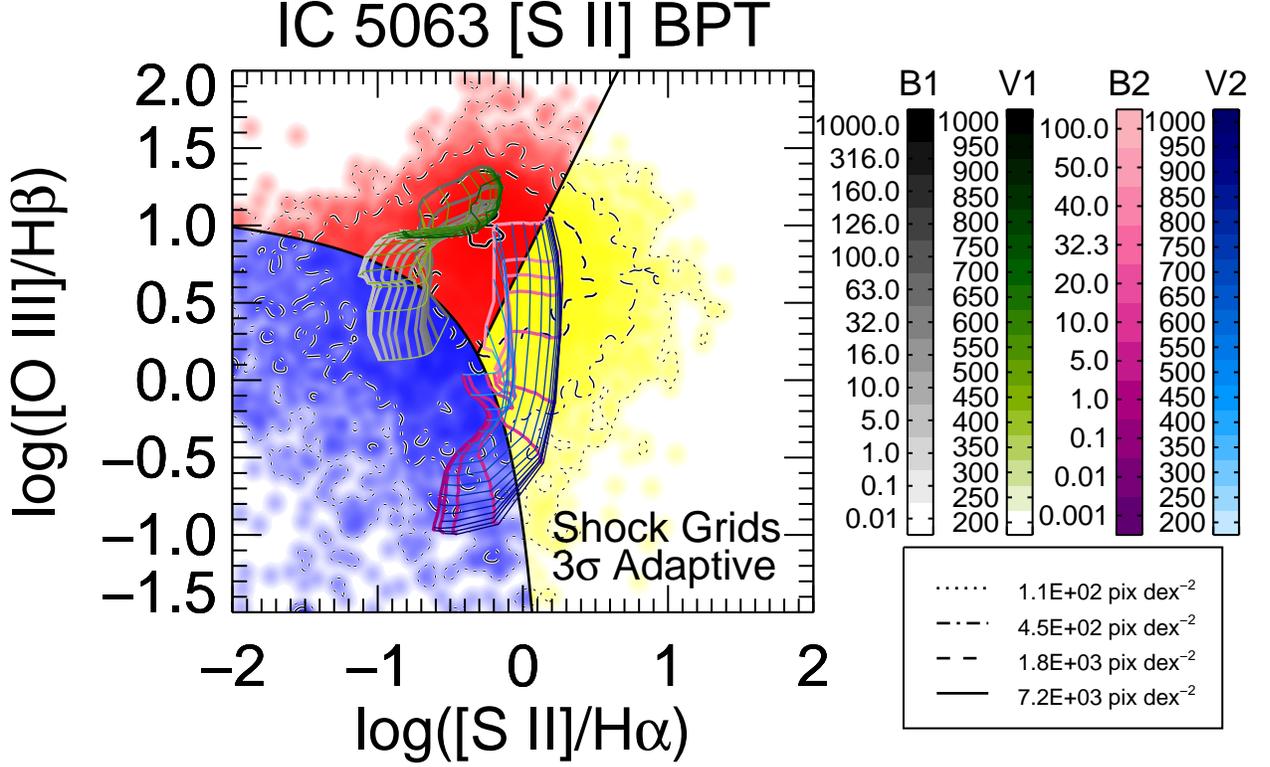}
\caption{As for Fig. \ref{fig:bpt} (lower right), but with \citep{Allen08} MAPPINGS\,III shock model grids overlaid.  Model 1 (B1=magnetic field in $\mu$G, V1=velocity in $\kms$) assumes electron density $n_e=1000\,\rm{cm}^{-3}$ and includes emission from the shock-photo-ionized precursor material.  Model 2  (B2=magnetic field in $\mu$G, V2=velocity in $\kms$) assumes electron density $n_e=1000\,\rm{cm}^{-3}$ and uses shock emission only.  The $n_e=1000\,\rm{cm}^{-3}$ shock-only grids are excluded for clarity.  They are located in the phase space between Models 1 \& 2, with iso-velocity contours roughly parallel to those for Model 2.
}
\label{fig:bptshock}
\end{figure*}

Our BPT map can be directly compared against more recent results from  \cite{Mingozzi19} using MUSE.  As with our {\it HST} data \cite{Mingozzi19} find Seyfert-like emission along the brightest [\ion{O}{3}] that follows the NW-SE radio axis.  Our {\it HST} data are less sensitive but have better spatial resolution (undersampled at $\sim0\farcs09$ pixels with WFPC2 vs. $\sim0\farcs8$ seeing with MUSE), with the caveat that care must be taken in determining the amount of [\ion{N}{2}] contamination in H$\alpha$.  In general the median value of [\ion{N}{2}]/[\ion{S}{2}] taken from Fig. \ref{fig:musen2s2} should be reliable, but lower [\ion{N}{2}] contamination becomes important for fast shocks, and this becomes most relevant in the cross-cone, and is therefore also relevant for lateral outflows.

We identify LINER-like emission where the NW bicone emission bifurcates.  This is not identified by \cite{Mingozzi19} and may be washed out at the resolution of the MUSE data.  We also identify LINER-like emission in the SW cross-cone, similarly to \cite{Mingozzi19}.  {\it HST} resolution pushes the cross-cone LINER to $\simless20\,$pc of the nucleus in the SW.  Given the symmetry observed by \cite{Mingozzi19} and uncertainties in our NE continuum scaling (due to dust), LINER-like emission could also be present at similar scales in the NE.

The [\ion{S}{2}] loop that we find in the SW with {\it HST} is roughly co-spatial with their cross-cone LINER, but is only marginally resolved in a the MUSE datacube, appearing as a short radial feature.   The typical line ratios we find in the loop ([\ion{S}{2}]/H$\alpha\simgreat 1.3$, [\ion{O}{3}]$\simless1\sigma$/pixel) are consistent with the  \cite{Mingozzi19} LINER, suggesting either shock excitation (as might be found in a lateral outflow), or EUV shielding from the AGN by intervening clumpy material on scales of $\sim10$s of pc \citep{Alexander00,Kraemer08,Netzer15,Mingozzi19}.

Although our ratio from the loop's integrated emission is LINER-like (\S\ref{sec:loop}), the BPT maps in Figs. \ref{fig:bpt}, \ref{fig:bpthi}, \ref{fig:bptlo} show a patchy mix of structure that includes mainly LINER-like and SF pixels with some Seyfert-like activity.  The patchy pixel-to-pixel variation outside the jet and bicone could imply (if real) that unresolved measurements of extended ionization structure may be dominated by emission from small ($\sim20-40$\,pc; $\sim1-2$ $0\farcs09$ pixels across), bright clumps.  SNR fluctuations could be important for producing this effect, but the major line requiring adaptive smoothing is H$\beta$, whereas the most relevant diagnostic observations (H$\alpha$ and [\ion{S}{2}]) are generally more sensitive.  The PSF is under-sampled at the pixel scale ($D\sim20$\,pc) in these ratio maps, so single-pixel features cannot automatically be attributed to noise.

Although it is not possible to unambiguously distinguish between shocks and photoionization with only these lines, we can discriminate between specific conditions within a given scenario.  \cite{Mingozzi19} infer electron densities $n_e$ of $10^2\lesssim n_e\lesssim10^3\,\rm{cm}^{-3}$ for [\ion{S}{2}]-emitting gas in the nucleus. In Fig. \ref{fig:bptshock}, we overplot the BPT diagram in Fig. \ref{fig:bpt} with two grids of shock models from \cite{Allen08} assuming \cite{Mingozzi19} MUSE-derived densities, comparing models with pre-shock $n_e=10^3\,\rm{cm}^{-3}$ and solar abundance that incorporate a shock-photoionized precursor against models with no precursor and $n_e=10^2\,\rm{cm}^{-2}$.  We also examined models with $n_e=10^3\,\rm{cm}^{-3}$ and no precursor, as well as  $n_e=10^2\,\rm{cm}^{-3}$ and precursor.  Shocks with a precursor can explain SF-like emission (with $v_s\lesssim450\kms$) and much of the Seyfert-like locus (with $v_s\simgreat500\kms$), but LINER-like emission requires shocks to have no precursor.  To be consistent with higher [\ion{S}{2}]/H$\alpha$ ratios and $2\sigma$ uncertainties of $\sim0.4$\,dex, $n_e=10^2\,\rm{cm}^{-3}$ and higher velocities approaching $\sim1000\kms$ are favored.  But \citep[as in][]{Perna17,Mingozzi19} \cite{Allen08} fails to reproduce the largest [\ion{S}{2}]/H$\alpha$ values.  \cite{Mingozzi19} suggested that such a discrepancy could result from metallicity effects.

\subsection{Lateral Bubbles, Plumes and Outflows}

\cite{Mukherjee18} invoke simulations of inclined jets to describe IC 5063, and these simulations show plumes of ablated material carried by hot bubbles that vent perpendicularly to the disk.  The [\ion{S}{2}] loop could trace partially ionized gas enclosing just such a bubble.  The bubble could be analogous to those found in Circinus \citep{VBH97}.  If the bubble is generated by lateral outflows as per  \cite{Mukherjee18}, then there is no need here to invoke black hole mode-switching, as has been proposed for the {\it voorwerpje} bubbles \citep{Keel15,Keel17,Sartori16}.
Some of the sub-arcsecond loops identified by \cite{Keel17} might also be driven by lateral outflows, such as for SDSS J220141.64+115124.3.

The \cite{Mukherjee18} simulations may also explain the need for precursorless emission for LINER features which are shock-driven, as well as more extreme [\ion{S}{2}]/H$\alpha$ ratios.  In these simulations, jet-ISM interactions cause hot gas to vent perpendicularly to the disk, filling the extraplanar ISM with $n_e\lesssim1\rm{cm}^{-3}$, $T\sim10^9$\,K gas, cooler $T\sim10^7$\,K filaments, and smaller $T\sim10^5$\,K clumps.  \cite{Maksym20} describe $\sim11$\,kpc-scale dark features in the optical/NIR continuum emission as ``dark rays" because they extend radially from the nucleus of IC 5063, nearly orthogonally to the jet and disk.

If these ``dark rays" result from kpc-scale hot outflows removing reflective dust from the local path of AGN radiation (as possibly supported by [\ion{O}{3}] line width maps, \citealt{Venturi21}), then such hot outflows could extend  well beyond the extraplanar regions explored in this paper. The pre-shock gas would therefore have very different properties than typically assumed in ISM shock models. \cite{Allen08} assume a discontinuous temperature jump between the ionization region ($\sim10^6$K) and the precursor ($\lesssim10^4$K).  But in IC 5063, the bulk of the pre-shock gas could be quite hot ($\simgreat10^6$K) and hence unable to emit the strong [\ion{O}{3}]  that characterizes fast shocks, as well the radiatively excited pre-shock Balmer emission \citep[see][for examples of ionic species stratification structure in shocks]{DS96}.  Suppose the speed of sound is  $c_s=\sqrt{\gamma k_B T/(\mu m_p)}$, where $\gamma=5/3$ is the adiabatic index for a monoatomic gas, $k_B$ is the Boltzmann constant, T is the temperature, and $\mu m_p=0.59 m_p$ is the average particle mass in terms of the proton mass $m_p$ \citep{Zinger16}.  LINER-like shocks should then be possible for the optical line velocities observed in IC 5063 ($v\simgreat 500\kms$) even against hot ($T\simgreat10^7$\,K) filaments in the bulk outflow.  Such shocks would be only modestly supersonic (Mach $\sim1-2$).

In the case of IC 5063, the volume of such a bubble can constrain the total energy needed for an outflow to inflate it. Suppose bubble energy $E_{bubble}=(3/2)n_e V kT$, where $n_e$ is the electron density of the hot interior gas, $V$ is bubble volume, $k$ is the Boltzmann constant and $T$ is the interior temperature.  Then $E_{bubble}=1.2\times10^{55}(n_e/{\rm cm}^{-3})(T/10^{9}{\rm\, K})\,{\rm erg}$  Assuming an expansion velocity $v$ and major axis $A_{bubble}$, we infer mean kinematic luminosity 

\begin{equation}
\begin{split}
L_{KE,AGN}=&\frac{E_{bubble}v}{A_{bubble}}= \{3.0\times10^{41}\es \times\\
		&\frac{n_e}{{\rm cm}^{-3}}\frac{T}{10^{9}{\rm\, K}}\frac{v}{500\,\kms}\}
 \end{split}	
\end{equation}

\noindent is required for the AGN to inflate the bubble, which is $\sim0.04\%$ of $\Lbol$ and $\lesssim0.6\%$ of the total jet $L_{KE,jet}$ inferred by \cite{Mukherjee18} (\S4.2).  We can compare this to the supernova kinematic luminosity such that

\begin{equation}
\begin{split}
L_{KE,SN}= & \{6.4\times10^{40}{\es}\times \\
&\quad \frac{R_{SF}}{5\Msun{\rm yr}^{-1}}\frac{\eta}{10^{-3}}\frac{KE_{SN}}{10^{51}\,{\rm erg}}\frac{f_{SF}}{0.04} \}
 \end{split}	
\end{equation}

\noindent where the galactic star formation rate is estimated by \cite{Mukherjee18} via FIR emission, $R_{SF}$ is converted to a supernova rate via an efficiency rate $\eta$, the amount of kinetic energy released to the ISM per supernova is $KE_{SN}$, and the fraction of total star formation in the nucleus is $f_{SF}$ (inferred by the ratio of F763M continuum emission within $r<1\arcsec$ vs. the whole galaxy).  We see that although $L_{KE,SN}$ could be significant, $L_{KE,AGN}$ clearly dominates for reasonable assumptions.  ${f_{SF}}$ is conservatively large here; the base of the bubble overlaps with a $r\simless0\farcs35$ region at the base of the jet.  The relatively well-defined elliptical shape of the bubble is also contrary to expectations for a more broadly distributed wind driven by episodic nuclear supernovae.

Fig. \ref{fig:linepics} shows evidence (particularly in [\ion{S}{2}]) of additional filamentary structure that suggests the circumnuclear ISM is ``frothy", such that this loop happens to be the brightest and best-defined of several such structures.  This structure is indicated by yellow circles and dotted lines in the H$\alpha$, [\ion{S}{2}] panel of Fig. \ref{fig:linepics}. Circumnuclear supernovae might also contribute to such bubbles, but their impact would be difficult to disentangle from the known bright AGN.  

The aggregate effects of this ``frothy" outflow structure may even be seen at larger scales, as MeerKat observations by \cite{Morganti98} show radio structure extended on $\sim15\arcsec$ scales in this direction.  Displacement of dust by such large-scale outflow structure could be responsible for displacing diffuse reflective dust and thereby causing the dark ``rays" described by \cite{Maksym20}.  
High resolution X-ray, radio and optical observations show that such multiphase outflows can be stratified, with discrete and complementary structures \citep[e.g.][]{Maksym17}, and suggest that the outflows may produce extended regions of collisionally excited X-ray gas outside of the bicone and the sites of strong shocks \citep{Fabbiano18,Maksym19}.  The expected low density of the X-ray emitting gas implies that such emission requires very sensitive observations to detect, but the narrow line structure seen here and the dust displacement in IC5063 \citep{Maksym20} may provide alternate evidence for such phenomena.

The link between the accretion state of circumnuclear gas in an AGN and its accretion rate is obvious due to the need for energetic photons, and this is necessarily tied to the availability of fuel in the form of the nuclear ISM.  IC 5063 points to a possible role for the thermodynamic state of the ISM, such that less-ionized signatures may be produced in the absence of cool ISM, which prevents shock-driven photoionization.  To the extent that BH mass scales with bulge properties, the Eddington ratio could be dependent upon galaxy type and \Mbh\, while for a fixed gas mass, other aspects like the gas density, temperature and kinematics should depend on host properties \citep{Ho09a,Ho09b,YH20}.

\subsection{Azimuthally Stratified Photoionization}

\citet[Fig. 4a-c]{Maksym20} propose that an optically thin dusty halo extended on $\sim10$-kpc scales could reflect the AGN continuum emission in IC 5063 towards the observer, and that one possible cause of dark radial features seen on these spatial scales in the $\sim$I band continuum (F763M and F814W filters; Fig. 1c, 1d, 4b \citealt{Maksym20}, and as indicated in Fig. \ref{fig:linepics}a in this paper) might be shadows cast by the torus or other nuclear dust lane.  In this scenario, a broad opening angle ($\sim140\degr$) is required to produce such shadows with non-ionizing radiation.   Such a scenario would complement azimuthal stratification of ionizing radiation relative to the bicone axis, which may be indicated by the highly ionizing Seyfert-like regions in the bicone and the LINER zones immediately outside the bicone.  The Seyfert-like emission mainly tracks [\ion{O}{3}] in the bicone which subtends an angle of $\simless50\degr$ (compare Fig. \ref{fig:linepics} and Fig. \ref{fig:bpt}), although some of the strong [\ion{O}{3}] in the bicone may be stimulated by photoionizing shocks in jet-ISM interactions.  The transition from Seyfert-like to LINER-like and SF-like emission is rapid in the adaptively smoothed BPT maps (Figs. \ref{fig:bpt}, \ref{fig:bpthi}, \ref{fig:bptlo}), covering only $\sim20-80$\,pc in projection.

The [\ion{S}{2}] loop or bubble immediately SW of the nucleus (and within the putative shadow) is weak in [\ion{O}{3}] and LINER-like overall, implying a relative lack of EUV and X-rays intercepted from the nucleus.  This would be consistent with preferential obscuration by intervening material at large angles from the bicone axis, and suggests that the loop's [\ion{S}{2}] emission should be stimulated by shocks.

{\it HST} resolution is necessary to resolve the diagnostic line structures seen here, \cite{Maksym16} and \cite{Ma21}. Adaptive smoothing here points to low-ionization SF-like emission outside the nucleus, which is a necessary constraint on the subtended angle for any LINER transition zone at the edge of the bicone, or shocks outside the bicone produced by lateral outflows.   This is evident from \cite{Ma21}, the first comprehensive diagnostic line survey of AGN at {\it HST} resolution, which similarly identifies numerous LINER cocoons and ionization morphology which may be due to complex outflows. But \cite{Ma21} do not adaptively smooth and generally fail to map significant diagnostic ratios either in the SF-like region of the BPT diagram or $\simgreat1\arcsec$ into the cross-cone.  Deeper surveys (particularly ones sensitive to faint H$\beta$) would therefore improve our understanding in this transition, and more generally how the geometry of active nuclei affects observed ionization signatures.  Maps of additional line species could also help break the challenging degeneracies between photoionization and shocks intrinsic to this kind of study.

\section{Conclusion}

Via BPT mapping achieved via {\it HST} narrow filter imaging, we confirm that gas in the bicone (which includes the sites of strong jet-ISM interactions) is predominantly Seyfert-like $\simless1.4$\,kpc from the nucleus.  On {\it HST} pixel scales \citep[too small for MUSE to resolve in][]{Mingozzi19} there is evidence for a sharp transition to LINER ratios outside the bicone, similarly to NGC 3393 in \cite{Maksym16}.  This points to possible angular stratification of the incident ionizing spectrum, relative to the bicone axis and suggests a sharp transition to lower-ionization states outside the jet path.  Such emission is likely dominated by $\sim10-40$\,pc clumps and filamentary structure at large ($>>25\degr$) angles from the bicone axis.

In the BPT bands examined ([\ion{S}{2}], [\ion{O}{3}], H$\alpha$, H$\beta$), we find a $\simgreat700$\,pc-scale loop perpendicular to the bicone and within the dark ``rays" previously examined by \cite{Maksym20}.  This loop may be poorly resolved by MUSE and with {\it HST} is only reliably identifiable in [\ion{S}{2}] (given significant likely [\ion{N}{2}] contamination in the H$\alpha$ observations).  We expect clumps in this loop to comprise the major component of the LINER structure identified by \cite{Mingozzi19}, and for local MUSE-derived [\ion{N}{2}]/[\ion{S}{2}] ratios (appropriate to   fast $\sim1000\,\kms$ shocks) the loop region is indeed LINER-like.  Under our assumed geometry, the loop could result from a bubble of hot gas driven by jet-ISM interactions, ablating local material and rising laterally from the galactic plane, as in \cite{Mukherjee18}.  Bubbles and plumes of gas driven by such processes may stimulate LINER-like ionization via precursorless shocks.  They may also combine to form a larger lateral outflow that can entrain, displace or destroy dust at scales of $\sim10$s of kpc, and are a possible explanation for the dark ``rays" described by \cite{Maksym20}.

Our new deep {\it Chandra} observations of IC 5063 (PI: Fabbiano) should have sufficient spatial resolution to investigate the role of hot outflowing gas in producing such plumes.  If low-ionization lateral outflows (such as may cause the loop in IC 5063) are common in AGN, further high-resolution narrow-line observations are necessary to reveal them, especially in [\ion{S}{2}].  Deep {\it HST} continuum observations at blue and ultraviolet wavelengths could help reveal specific feedback effects of lateral outflows on star formation in IC 5063 by determining the absence or presence and location of young stellar populations.

\vspace{2in}

\acknowledgments

Special thanks to Raffaella Morganti for helping to develop the joint {\it Chandra/HST} program that led to this paper, and to both Raffaella Morganti and John Raymond for helpful scientific discussions.

WPM acknowledges support by Chandra grants GO8-19096X, GO5-16101X, GO7-18112X, GO8-19099X, and Hubble grant HST-GO-15350.001-A.  This work was also supported by Hubble grant HST-GO-15609.001-A from the Space Telescope Science Institute, which is operated by AURA, Inc., under NASA contract NAS 5-26555.

 LCH was supported by the National Science Foundation of China (11721303, 11991052) and the National Key R\&D Program of China (2016YFA0400702).

We thank the referee for many useful comments that greatly improved the quality of the paper.

%

\facilities{HST(WFC3, WFPC2)}







\end{document}